# Vers une ontologie de domaine pour l'analyse des tissus anciens

## Le projet SILKNOW et le cas du patrimoine soyeux européen

# Towards a Domain Ontology for the Analysis of Ancient Fabrics

## The SILKNOW Project and the Case of European Silk Heritage

Marie Puren[1] et Pierre Vernus[2]

**Résumé en français et en anglais**

Dans cet article, nous présentons le projet SILKNOW (*Silk heritage in the Knowledge Society: from punched card to Big Data, Deep Learning and visual/tangible simulations*) (2018-2021). Ce projet avait pour but d'utiliser les technologies du Web sémantique pour valoriser des objets en soie produits et consommés en Europe entre le XVème siècle et le XIXème siècle. La soie est un matériau particulièrement important dans l'histoire européenne, et elle a permis de produire des objets exceptionnels et d'un grand intérêt historique. Il s'agit pourtant d'un patrimoine menacé et mal connu du grand public. Nous montrons l'intérêt d'utiliser les technologies du Web sémantique pour donner plus de visibilité à un tel patrimoine, en décrivant les résultats que nous avons obtenus. Nous exposons la méthodologie utilisée pour développer un graphe de connaissances, et tout particulièrement les différentes étapes qui ont été nécessaires pour la création du modèle de données sous-jacent, basé sur le modèle conceptuel CIDOC CRM ou CIDOC *Conceptual Reference Model*. Nous proposons également une extension compatible avec le CIDOC CRM, destinée à exprimer la sémantique complexe du processus de création et de production des tissus anciens en soie.

In this article, we present the SILKNOW project (*Silk heritage in the Knowledge Society: from punched card to Big Data, Deep Learning and visual/tangible simulations*) (2018-2021). This project aimed to use Semantic Web technologies to give greater visibility to silk objects produced and consumed in Europe between the 15th and 19th centuries. Silk is a particularly important material in European history, and it has produced some exceptional objects of great historical interest. However, it is a threatened heritage that is little known to the general public. We show the interest of using Semantic Web technologies to give more visibility to such a heritage, by describing the results we have obtained. We present the methodology used to develop a knowledge graph, and in particular the different steps that were necessary to create the underlying data model, based on the CIDOC CRM or CIDOC

[1] ORCID : 0000-0001-5452-3913. Rattachements : MNSHS, Epitech, Kremlin-Bicêtre et Centre Jean-Mabillon, Ecole nationale des chartes, Paris. Adresse mail : marie.puren@epitech.eu. Marie Puren est enseignante-chercheuse en histoire et humanités numériques.

[2] ORCID : 0000-0002-9335-7070. Rattachements : LARHRA, CRNS et Université Lumière Lyon 2, Lyon. Adresse mail : Pierre.Vernus@ish-lyon.cnrs.fr. Pierre Vernus est maître de conférences en histoire contemporaine.

Conceptual Reference Model. We also propose a CIDOC CRM-compatible extension to express the complex semantics of the creation and production process of ancient silk fabrics.

# Table des illustrations

# Introduction

La croissance du Web a mené à une redéfinition de la stratégie en matière de diffusion des données patrimoniales. Si les institutions patrimoniales[3] sont conscientes de l'importance d'avoir une présence institutionnelle sur le Web pour séduire et fidéliser de nouveaux publics (Bourgeaux 2009) (Gomes 2017), elles ont également saisi l'intérêt de diffuser leurs données en ligne, en vue de bénéficier des nouvelles possibilités offertes par le Web sémantique pour l'exposition et la valorisation de ces données (Domange, Guide Data Culture. Pour une stratégie numérique de diffusion et de réutilisation des données publiques numériques du secteur culturel 2013) (Chaudiron, Jacquemin et Kergosien 2018). La numérisation du patrimoine, et la valorisation des données numériques qui en résulte, sont donc devenus des objectifs cruciaux pour les établissements de conservation (Eastermann 2015). On voit même apparaître un nouveau paradigme, selon lequel on doit traiter les collections (numériques) comme des données (« *collections as data* »), et pas seulement comme des collections d'objets patrimoniaux numérisés. Ce type de conceptualisation amène à déployer de nouvelles approches, notamment à développer des méthodes pour faciliter la découverte et la réutilisation de ces données, et à penser les métadonnées comme des données, au même titre que les collections numériques (Ziegler 2020). Pour accroître la découvrabilité de ces données, on mise aujourd'hui de plus en plus sur les possibilités offertes par le Web sémantique.

Les perspectives ouvertes par le Web sémantique sont particulièrement séduisantes pour les institutions patrimoniales : celui-ci leur propose d'accroître la visibilité de leurs métadonnées sur le Web en les « libérant » des catalogues, d'agréger ces données avec d'autres et d'y donner accès de manière fédérée pour encourager leur réutilisation, d'interconnecter les données entre elles de façon à créer un vaste réseau d'informations, et *in fine* d'enrichir plus facilement ces informations grâce aux liens créées entre elles. Au vu des possibilités offertes par le Web sémantique pour la recherche et l'enrichissement des informations, la recherche historique peut grandement bénéficier de ces technologies. En encourageant la diffusion de données structurées, ces dernières facilitent notamment la création de liens entre des sources dispersées (Meroño-Peñuela, et al. 2015). La protection, la sauvegarde et la valorisation du patrimoine constituent donc des enjeux importants pour le Web sémantique (Chaudiron, Jacquemin et Kergosien 2018), et offrent un nouveau cas d'application de ces technologies. Les patrimoines menacés de disparition (ou « *Word Heritage in Danger* » au sens de l'UNESCO[4]) peuvent notamment tirer parti des possibilités d'agrégation, de valorisation et d'enrichissement des données (Chaudiron, Jacquemin et Kergosien 2018). C'est dans ce cadre que s'inscrivait le projet européen SILKNOW (*Silk heritage in the Knowledge Society : from punched card to Big Data, Deep Learning and visual/tangible simulations*)[5], qui avait pour but d'utiliser les technologies du Web sémantique pour valoriser le patrimoine soyeux européen. Il s'agit en effet d'un patrimoine fragile, composé de collections dispersées dans de nombreux musées à travers le monde, et encore mal connu du grand public. SILKNOW a fait l'hypothèse qu'une meilleure exposition de ces collections assurerait une meilleure protection de ce patrimoine, en leur donnant plus de visibilité, en stimulant de nouvelles recherches sur ces dernières et en encourageant la réutilisation des métadonnées les décrivant. Nous nous intéressons en effet aux métadonnées descriptives, « qui servent à organiser la connaissance […] et qui vont permettre d'identifier, classifier, hiérarchiser l'information contenue dans l'objet numérique[6] ». Entre 2018 et 2021, le projet a ainsi développé et mis en ligne un moteur de recherche exploratoire intitulé ADASilk, qui permet de réaliser une recherche fédérée au sein de collections patrimoniales consacrées à la soie.

[3] Dans le cadre de cet article, les institutions patrimoniales rassemblent les musées, les archives et les bibliothèques.
[4] Pour la liste des « *Word Heritage in Danger* » dressée par l'UNESCO : http://whc.unesco.org/en/danger/
[5] Le site du projet est accessible à cette adresse : https://silknow.eu/

ADASilk repose sur un *Knowledge Graph* ou « graphe de connaissances » qui permet d'agréger et de présenter en ligne des données provenant de sources variées. Le SILKNOW *Knowledge Graph* représente des données issues des catalogues de vingt et une institutions patrimoniales[7], en suivant un sous-ensemble du modèle conceptuel CIDOC *Conceptual Reference Model* ou CIDOC CRM. SILKNOW a, la plupart du temps, collecté ces données en ligne ; seules trois institutions[8] ont directement fourni les jeux de données qu'elles souhaitaient voir intégrées au *Knowledge Graph*. Toutes ces données décrivent des objets intégrant de la soie dans leur processus de fabrication, et qui ont été produits ou consommés en Europe entre le milieu du XVème siècle et le milieu du XIXème siècle. Ces objets peuvent être totalement en soie, ou bien incorporer d'autres fibres (fibres naturelles comme la laine, le coton, etc., fils en métal comme l'or et l'argent), ou encore comporter des matériaux composites (par exemple, le bois d'un meuble). Les objets décrits sont particulièrement variés, comme le montrent les figures 1, 2, 3, 4 et 5 : des pièces d'habillement (civils et religieux), des accessoires, des fragments ou pièces de tissu, ou encore des meubles.

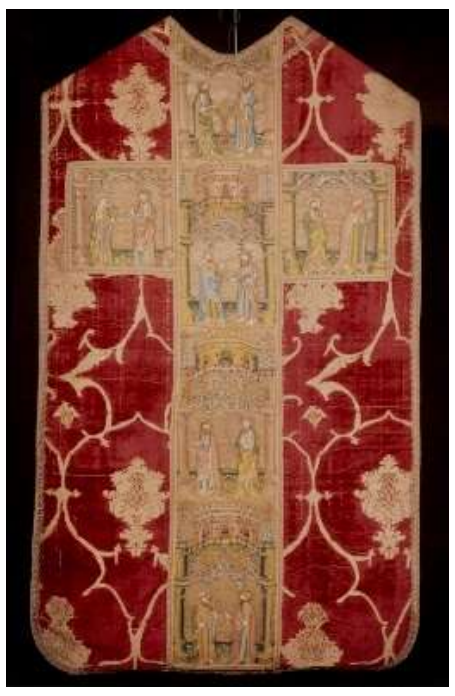

Figure 1. Chasuble en velours (Italie, 1426-1475) en soie et fils métalliques conservé au *Victoria & Albert Museum* (Londres) (©Victoria and Albert Museum, London)

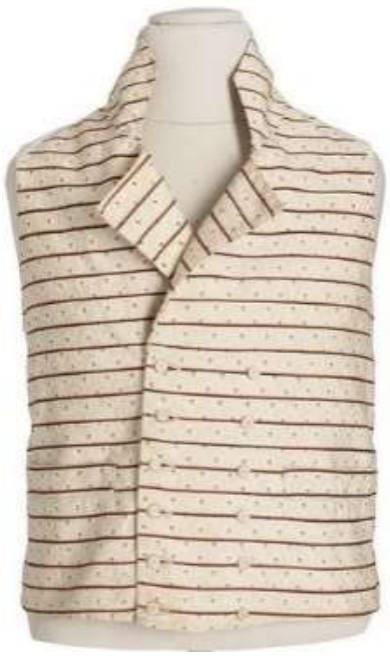

Figure 2. Gilet (France, 1795-1800) en soie et laine conservé au *Musée des Arts décoratifs* (Paris) (©Musée des Arts décoratifs, Paris)

---

[6] Nous utilisons ici la définition donnée par le Portail International Archivistique francophone : https://www.piaf-archives.org/sites/default/files/bulk_media/m07s09/co/Module_sections9_8.html
[7] La liste des institutions peut être consultée à cette adresse : https://ada.silknow.org/fr/museums
[8] Il s'agit de *Garin1820*, le *Sicily Cultural Heritage* et le *Musée d'Art et d'Industrie de Saint-Etienne.*

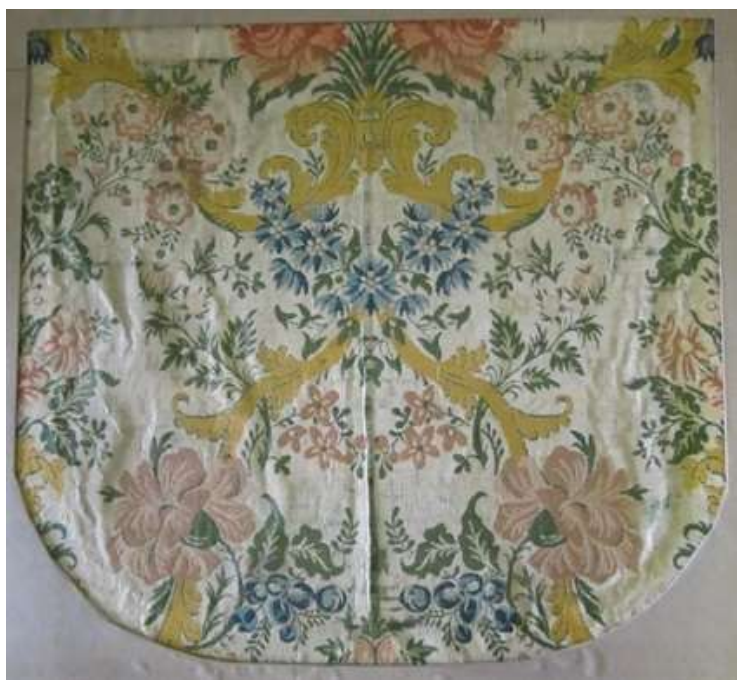

Figure 3. Tissu (Italie, XVIIIème siècle) en soie conservé au *Centro Studi di Storia del Tessuto, del Costume et del Profumo* (Venise) (©Centro Studi di Storia del Tessuto, del Costume et del Profumo, Venice)

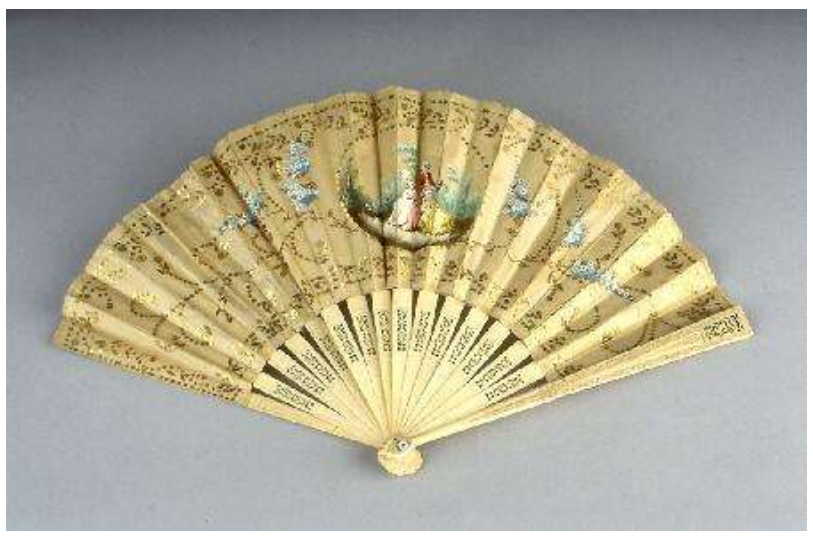

Figure 4. Eventail (Inconnu, 1775-1825) en nacre, ivoire et métal conservé au *Museo Nacional de Escultura* (Valladolid) (©Museo Nacional de Escultura, Valladolid)

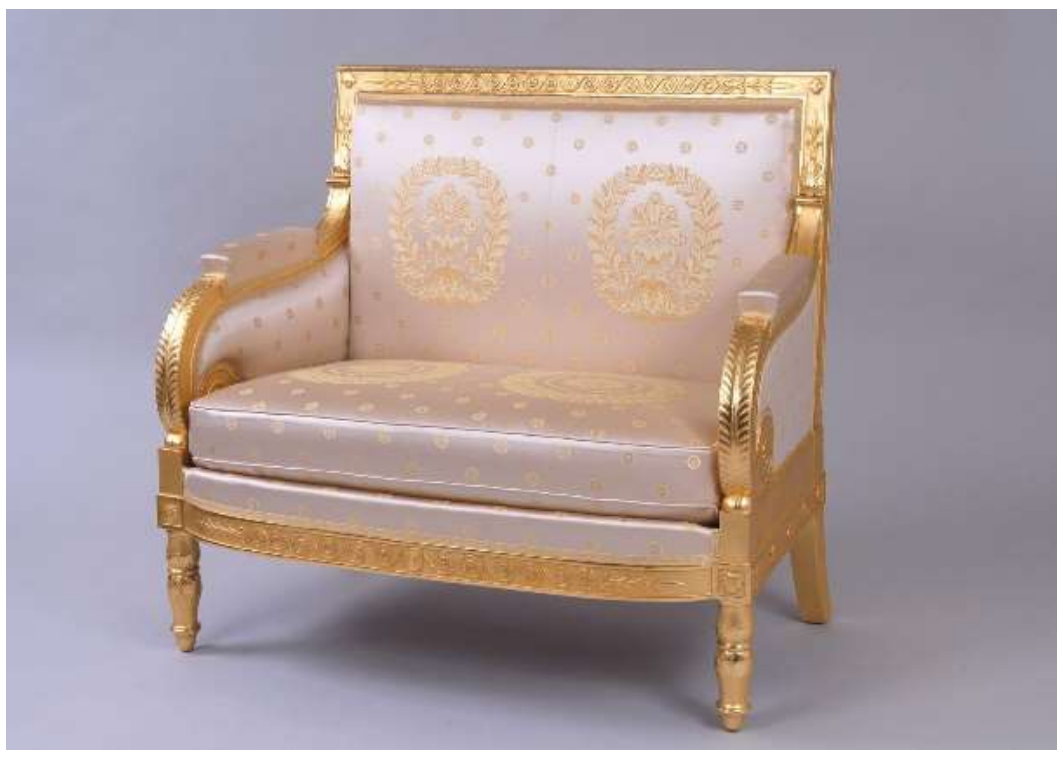

Figure 5. Causeuse (France, 1804-1815) en soie, or et bois conservé par le *Mobilier national* (Paris) (©Mobilier national, Paris)

Dans cet article, nous nous intéressons aux différentes étapes du travail qui ont abouti à la création du modèle de données SILKNOW sur lequel repose le graphe de connaissances SILKNOW. Dans une première partie, nous décrivons les enjeux entourant la protection du patrimoine textile, et les apports du Web sémantique pour sa conservation. Nous montrons que les métadonnées décrivant ce patrimoine sont particulièrement hétérogènes, et nous décrivons quelle stratégie SILKNOW a adoptée pour surmonter cette hétérogénéité et diffuser ces données en ligne. Dans une deuxième partie, nous exposons la méthodologie que nous avons utilisée pour créer le modèle de données SILKNOW basé sur le modèle de référence conceptuel CIDOC CRM. Dans une troisième partie, nous formulons des propositions pour intégrer, dans le modèle de données, des annotations issues de l'analyse des textes et des images, et pour exprimer la sémantique complexe des données décrivant la structure des tissus anciens en soie. Nous présentons notamment les nouvelles classes et propriétés que nous avons créées pour modéliser ces données.

# Mettre le Web sémantique au service du patrimoine soyeux

## Patrimoine textile, patrimoine soyeux sur le Web : bilan et enjeux

Depuis une dizaine d'années, la mode s'expose de plus en plus dans les musées, car elle attire un public croissant (Petrov 2019). Elle est devenue un véritable « produit d'appel » pour ces institutions, qui espèrent ainsi augmenter le nombre de visiteurs et séduire de nouveaux mécènes (Stewart et Marcketti 2012). Depuis une vingtaine d'années, on a redécouvert le patrimoine textile grâce à un regain des travaux à son sujet (ACAOAF 2008). Mais malgré cet intérêt renouvelé, il s'agit d'un patrimoine menacé de disparition. C'est d'autant plus le cas du patrimoine soyeux dont la grande fragilité pose d'importantes difficultés de conservation. La protection des tissus en soie, par nature extrêmement fragiles, demande de réaliser des investissements financiers importants et d'avoir recours à des experts du tissage en cas de restauration ou de reproduction. Car c'est aussi un patrimoine culturel immatériel[9], constitué (entre autres) par des pratiques et techniques de tissage (Chaudiron, Jacquemin et Kergosien 2018) (Kergosien, Severo et Chaudiron 2015), elles aussi menacées de disparition. Peu de matériaux ont pourtant comme la soie une telle importance historique, culturelle et artistique. La soie a été abondamment utilisée – et est encore utilisée - dans toute l'Europe pour produire des costumes, du mobilier ou encore des accessoires, pendant plusieurs siècles et selon des traditions variées. On trouve ainsi de nombreux tissus anciens dans les collections des grands musées ; mais paradoxalement, une bonne partie du patrimoine soyeux est conservé par des institutions petites en taille, qui manquent généralement de moyens pour valoriser leurs collections (Portalés, et al. 2018) (Gaitán, et al. 2019) (Pagán, et al. 2020).

Si la mode et les objets textiles prennent une place croissante dans l'espace muséal, cela n'a toutefois pas un impact direct sur l'utilisation des collections textiles par les usagers. L'intérêt pour ces collections se limite en général aux grands événements qui leur sont consacrés ; et en moyenne, on constate une sous-utilisation de ces collections par les usagers (Stewart et Marcketti 2012). Pour mieux les valoriser, les institutions patrimoniales misent en grande partie sur Internet, et notamment sur la mise en ligne de leurs inventaires. Malheureusement, ces métadonnées ne sont pas facilement accessibles, car les catalogues qui les contiennent font partie de ce « Web profond » (ou « *deep Web* ») qui n'est pas indexé par les moteurs de recherche. Ces données seront donc utilisées par celles et ceux qui savent déjà que ces catalogues existent. Les catalogues agissent en fait comme des « silos de données » qui sont incompatibles entre eux, et donc mutuellement incompréhensibles (Halevy 2005). Il faut donc « libérer les données » des catalogues et des bases de données dans lesquelles elles sont enfermées pour mieux les faire connaître (Bermès 2013).

Pour donner plus de visibilité à ces données et en faciliter la réutilisation, une solution consiste à créer un portail qui agrège des sources hétérogènes et y donne accès de manière simultanée. Cette solution est de plus en plus privilégiée par les institutions patrimoniales qui, dans un contexte de crise économique, souhaitent nouer de nouveaux partenariats pour mettre en place des projets de valorisation à grande échelle (Suls 2017). Les portails permettent ainsi à leurs utilisateurs de consulter les inventaires de collections qui sont, physiquement, conservées par de multiples institutions. On pourra par exemple citer le portail *Paris Musées*[10] consacré aux collections des musées de Paris, ou encore le portail des collections des Fonds régionaux d'art contemporain (FRAC)[11]. Du côté du patrimoine textile et de la mode, le portail *Europeana Fashion*[12] donne accès aux inventaires d'un peu plus d'une trentaine de musées. Mais ce portail se concentre avant tout sur les collections des grands musées européens, laissant de côté les petites institutions (Suls 2017). Par ailleurs il est difficile, avec ce type de solution, de mettre en avant des collections plus spécialisées, comme celles consacrées à la soie.

---

[9] Au sens de l'UNESCO : https://ich.unesco.org/fr/qu-est-ce-que-le-patrimoine-culturel-immatriel-00003
[10] https://www.parismuseescollections.paris.fr/fr

La fréquentation des sites Internet des musées et la consultation de leurs inventaires constituent un autre « porte d'entrée » pour découvrir le patrimoine textile. Les utilisateurs sont alors confrontés à deux cas de figures. D'un côté, les grands musées généralistes ont les moyens techniques, financiers et humains pour mettre en ligne leurs inventaires. Toutefois, même s'ils conservent de riches collections textiles, les musées ne mettent pas systématiquement en avant les objets appartenant à ces dernières. C'est le cas par exemple du *Victora & Albert Museum*[13] à Londres qui possède un important département consacré à l'histoire de la mode (Bide 2021), mais qui est aussi un grand musée d'art et de design. À contrario, les petits musées manquent souvent de moyens pour mettre en ligne leurs inventaires et maintenir leurs catalogues (Claerr et Westeel 2010). Il reste en effet du chemin à faire en matière de numérisation du patrimoine, car les collections patrimoniales sont loin d'être toutes numérisées et mises en ligne[14]. C'est notamment le cas des institutions qui se concentrent sur les collections très spécialisées - parfois particulièrement riches. Le patrimoine soyeux dans la région Rhône-Alpes est une bonne illustration de ce phénomène (Foron-Dauphin et Cano 2016) (Fournier, et al. 2016). Cela ne signifie pas que ces institutions ne décrivent pas leurs collections avec des métadonnées numériques ; mais ces données sont difficilement accessibles. Elles restent donc méconnues, sauf de quelques connaisseurs et spécialistes, selon le nouvel adage « ce qui n'est pas en ligne n'existe pas » (Ossenbach 2015).

# Protéger et valoriser le patrimoine avec le Web sémantique : quelques exemples

Grâce aux catalogues en ligne, les utilisateurs ont accès à distance aux métadonnées qui décrivent des objets conservés au sein de collections patrimoniales, que ceux-ci soient ou non exposés au public. Dans l'idéal, ces utilisateurs devraient pouvoir consulter plusieurs inventaires simultanément. Cela leur permettrait à la fois de resituer un objet patrimonial dans un contexte particulier (géographique, culturel, historique), et de le mettre en relation avec d'autres objets - que ceux-ci soient ou non conservés par le même musée (Le Boeuf, Le modèle conceptuel de référence du CIDOC : de la sémantique des inventaires aux musées en dialogue 2013). Dans les faits, les utilisateurs sont généralement confrontés à une multitude de catalogues, reposant sur des systèmes de gestion de l'information différents. Dans le cas d'un utilisateur qui s'intéresserait à des objets conservés dans des collections éparpillées à travers le monde, cela signifie qu'il doit se familiariser avec de nombreuses interfaces différentes. De plus, l'hétérogénéité de ces catalogues est encore accrue par la variété des langues utilisées, que les utilisateurs sont, implicitement, censés maîtriser. Par exemple, le catalogue collectif des musées espagnols ou *Red Digital de Colleciones de Museos de España* n'est disponible qu'en espagnol[15] ; il en va de même pour *Joconde*, le catalogue collectif des musées de France, qui n'est disponible qu'en français[16].

---

[11] http://www.lescollectionsdesfrac.fr/
[12] https://www.europeana.eu/fr/collections/topic/55-fashion
[13] https://www.vam.ac.uk/
[14] D'après le site *data.europa.eu*, 10% du patrimoine mondial était numérisé en 2018. Un peu plus du tiers était mis en ligne et 3% seulement sous une licence ouverte (https://data.europa.eu/en/highlights/cultural-institutions-and-cultural-open-data).
[15] http://ceres.mcu.es/
[16] Accessible via *POP : la plateforme ouverte du patrimoine* : https://www.pop.culture.gouv.fr/.

Le Web sémantique offre aujourd'hui de nouvelles perspectives de valorisation des données patrimoniales. Il ne s'agit pas d'un nouveau Web qui remplacerait le Web actuel ; il faut plutôt y voir une extension du Web tel que nous le connaissons aujourd'hui (Berners-Lee, Hendler et Lassila, The Semantic Web 2001). Grâce au Web sémantique, on ne naviguera plus entre les documents via des liens hypertextuels, mais directement entre les données : les données ne seront plus isolées les unes des autres, et le Web deviendra *in fine* une immense base de données. Ce qui implique qu'au préalable, toute information soit fournie dans un sens bien défini. C'est en effet ce qui va permettre de la rendre interprétable et réutilisable par les consommateurs de données, humains mais aussi machines qui seront alors capables de traiter automatiquement l'information, et de développer de nouvelles applications et de nouveaux services (Bermès 2013). Les institutions patrimoniales sont tout particulièrement intéressées par les possibilités d'agrégation offertes par les technologies du Web sémantique (Freire, Meijers, et al. 2018) (Freire, Voorburg, et al. 2019), notamment pour les métadonnées hétérogènes (Peroni, Tomasi et Vitali 2013).

C'est dans ce cadre que SILKNOW a fait l'hypothèse que la création d'un *Knowledge Graph* ou « graphe de connaissances » basé sur une ontologie, permettrait de faciliter l'intégration, l'exploration et l'extraction d'informations portant sur des objets et textiles anciens en soie. Un graphe de connaissances est un ensemble de descriptions interconnectées d'entités – objets, personnes, événements, concepts ou situations – qui peuvent être enrichies par des informations acquises à postériori[17] et être intégrées à des ontologies (Gruber 1993). Dans ce cas, on peut appliquer un raisonneur sémantique à ces données, et en inférer de nouvelles connaissances. Si les graphes de connaissances les plus connus sont certainement DBpedia[18] et Wikidata[19] (Carriero, Asprino et Biswas 2020), c'est aussi une approche qui commence à faire ses preuves en matière de valorisation des métadonnées patrimoniales hétérogènes ; celle-ci permet en effet de créer un point d'accès unique à ces données et d'en faciliter l'exploration. La multiplication des projets misant sur le développement d'un graphe de connaissances montre également la solidité de cette méthode. C'est grâce à un *Knowledge Graph* que le projet DOREMUS a, par exemple, développé un moteur de recherche permettant d'avoir accès à des données décrivant des œuvres musicales conservées par diverses institutions françaises (Achichi, et al. 2018). Récemment, l'utilisation de graphes de connaissances semble s'imposer comme l'une des solutions privilégiées lorsqu'il s'agit d'exposer des données patrimoniales hétérogènes. On peut ainsi citer le projet COURAGE (Faraj et Micsik 2021) et WarSampo (Koho, et al. 2019) qui traitent des données historiques variées, le projet ArchOnto (Koch, Ribeiro et Lopes 2020) qui s'intéresse à des données archivistiques, ou encore un projet portant sur la Compagnie Néerlandaise des Indes orientales (Schouten, et al. 2021).

## Quelles données et quelles technologies dans le cadre du projet SILKNOW ?

Les données traitées par le projet SILKNOW proviennent de vingt et une institutions patrimoniales. La plupart de ces données a été téléchargée en ligne, directement depuis les catalogues des institutions. Le projet SILKNOW a ainsi largement bénéficié de l'ouverture des données culturelles, qui s'inscrit plus largement dans le mouvement en faveur de l'« *open data* » ou « données ouvertes », que l'on définit comme des données « accessibles gratuitement, dans un format utilisable et modifiable pour servir tout objectif[20] ». L'ouverture des données est en effet le préalable indispensable à l'émergence du Web sémantique (Domange, Guide Data Culture. Pour une stratégie numérique de diffusion et de réutilisation des données publiques numériques du secteur culturel 2013). Les institutions culturelles se sont fait écho de ce mouvement en adaptant leur stratégie numérique en conséquence. Dans ce cadre, les consommateurs de données ne sont plus considérés comme des récepteurs « passifs », mais comme de possibles ré-utilisateurs à la recherche de données à exploiter (Domange, Ouverture et le partage des données publiques culturelles, pour une (r)évolution numérique dans le secteur culturel 2013). Cela se traduit à la fois par l'ouverture des reproductions numériques d'œuvres dans le domaine public, et par l'ouverture des métadonnées décrivant ces œuvres (Domange, Ouverture et le partage des données publiques culturelles, pour une (r)évolution numérique dans le secteur culturel 2013) (Jean, et al. 2012). Un certain nombre d'initiatives ont d'ailleurs fleuri pour favoriser la diffusion de ces données culturelles ouvertes. Par exemple, le réseau Open GLAM[21] rassemble des personnes et des organisations en faveur de l'ouverture des données produites et conservées par les « Galleries, Libraries, Archives and Museums » (GLAM). On peut également citer les initiatives *open data* portées par des institutions comme le *Rijksmuseum*, la *Digital Public Library of America* (Domange, Guide Data Culture. Pour une stratégie numérique de diffusion et de réutilisation des données publiques numériques du secteur culturel 2013), le *Carnegie Museum of Arts* ou encore le *Getty* (Ziegler 2020). Au niveau européen, la plateforme *Europeana* donne accès aux données numériques de plus de 3500 institutions européennes, en grande partie sous licence ouverte[22]. En France, la Bibliothèque nationale utilise une licence ouverte pour diffuser ses métadonnées bibliographiques depuis 2014[23] ; on peut également consulter la plateforme des données ouvertes du ministère de la Culture[24], ou la plateforme POP (Plateforme Ouverte du Patrimoine)[25] qui donne accès aux bases de données des Collections des Musées de France (base Joconde)[26].

[17] Par exemple, avec des annotations.
[18] https://www.dbpedia.org/
[19] https://www.wikidata.org/wiki/Wikidata:Main_Page

Le développement du Web sémantique a ainsi entraîné une mise en ligne accrue des données produites par les institutions patrimoniales (Eastermann 2015) (Marden, et al. 2013).Les musées, archives et bibliothèques sont d'importantes productrices de données numériques, dans des formats très divers et touchant tous les domaines du patrimoine (Domange, Guide Data Culture. Pour une stratégie numérique de diffusion et de réutilisation des données publiques numériques du secteur culturel 2013). Par nature, les métadonnées patrimoniales sont donc hétérogènes, car elles décrivent des objets de toutes sortes et sont produites dans des conditions d'une grande variété. Cela les rend d'ailleurs d'autant plus difficiles d'accès par les moteurs de recherche (Freire, Meijers, et al. 2018). Cette hétérogénéité est tout à fait visible dans les métadonnées collectées par SILKNOW - pas tant du point de vue du fond que de la forme. Les métadonnées sont composées de différents champs descriptifs, fournissant des informations identifiant l'objet numérique sans ambigüité (par exemple un identifiant unique, un titre, un auteur, une date, etc.). On utilise aussi l'expression « notices d'objet » pour désigner ces fichiers de métadonnées. Bien qu'elles proviennent de sources variées, les notices collectées par SILKNOW fournissent les mêmes types d'informations, généralement accompagnés d'une ou d'illustration(s) : le type d'objet, le lieu et la date de production, les matériaux et techniques utilisées, et les dimensions. Toutefois la structure de ces métadonnées varie en fonction de leurs producteurs. Les institutions qui les ont produites ont en effet pu adopter des normes de catalogage spécifiques, et faire évoluer leurs standards de description au fil du temps. Sur la figure 6, on voit que les deux notices, provenant respectivement du *Museu Tèxtil Terrassa*[27] et du *Victoria & Albert Museum*[28], offrent des informations similaires, mais qu'elles sont exprimées différemment. Par exemple, le *Victoria & Albert Museum* fournit, dans un même champ, des informations concernant le matériau et la technique utilisés, tandis que *Museu Tèxtil Terrassa* sépare ces informations.

---

[20] Nous reprenons ici la définition de l'*open data* dans Datactivist, *L'ouverture des données publiques culturelles en pratique*, Ministère de la Culture, sd., https://datactivist.coop/ministere-culture/jour2.html
[21] Pour en savoir plus sur ce réseau : https://openglam.org/
[22] https://www.europeana.eu/fr
[23] https://www.bnf.fr/fr/conditions-de-reutilisations-des-donnees-de-la-bnf
[24] https://data.culture.gouv.fr/
[25] https://www.pop.culture.gouv.fr/
[26] Pour d'autres exemples de projets *open data* dans les institutions patrimoniales, on pourra consulter le support de formation suivant : Datactivist, *Open Data : comprendre les enjeux de l'ouverture des données publiques culturelles*, Ministère de la Culture, https://datactivist.coop/ministere-culture/jour1.html#1

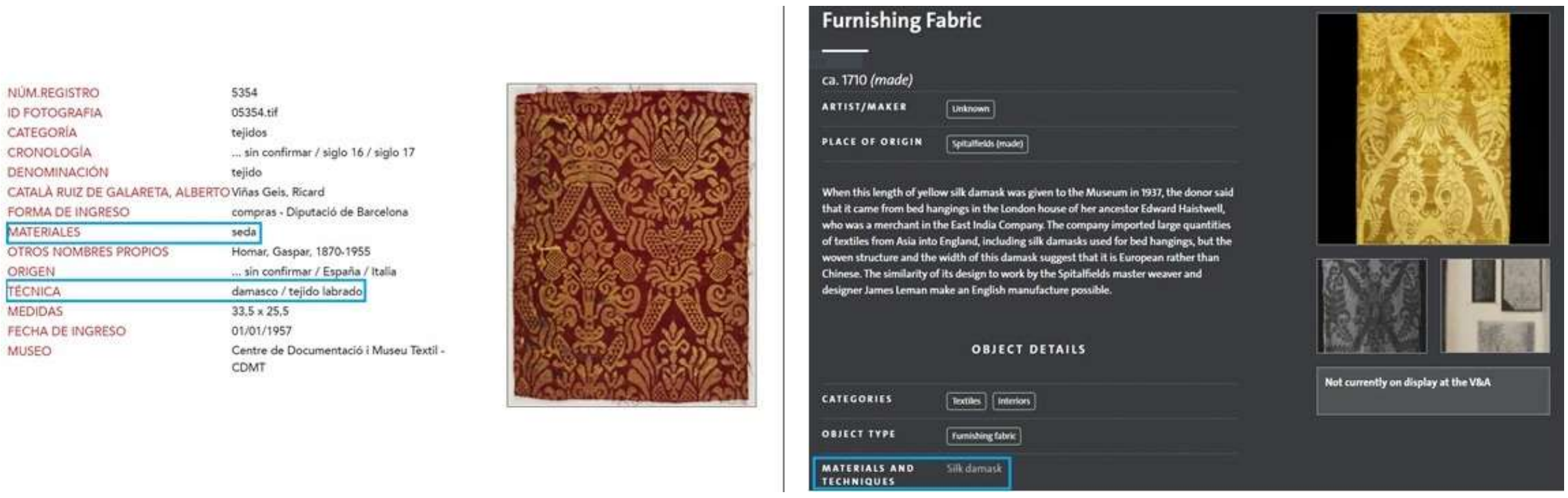

Figure 6. Notices décrivant de deux tissus « damas »[29] provenant du *Museu Tèxtil Terrassa* et du *Victoria & Albert Museum* (Image produite par les auteurs)

On se rappelle que ces catalogues se comportent comme des silos de données séparés, qui n'ont pas vocation à communiquer entre eux. L'hétérogénéité et le manque de structure des métadonnées collectées rendent également difficile leur traitement automatique en vue de leur exposition en ligne. Ce manque de structure, marquée notamment par la concaténation d'informations différentes dans un même champ (comme dans la figure 6), est l'un des principaux obstacles à leur analyse à grande échelle. Pour développer le réseau des données tel que l'imaginent les tenants du Web sémantique, cela implique donc de réaliser un important travail de préparation des données (Le Boeuf, Le modèle conceptuel de référence du CIDOC : de la sémantique des inventaires aux musées en dialogue 2013). Les données publiées par les institutions patrimoniales ne sont pas directement exploitables dans ce cadre, car elles ne peuvent pas être qualifiées de « données ouvertes liées ». Pour en faire des données interconnectées et lisibles par la machine, un ensemble de principes de conception a été proposé dès 2006 par Tim Berners-Lee (Berners-Lee, Linked Data 2006). L'application de ces principes permet de créer des données ouvertes cinq étoiles, c'est-à-dire (1) diffusées sur le Web, (2) mises à disposition sous forme de données structurées, (3) dans des formats ouverts, (4) identifiées par des URIs et (5) liées à d'autres données.

Après avoir collecté ces fichiers de métadonnées, le projet SILKNOW les a convertis pour créer le graphe de connaissances en utilisant le *Resource Description Framework* ou RDF (Schleider, Troncy et Ehrhart, et al. 2021) (Schleider, Troncy et Gaitan, et al. 2021). La mise en œuvre du Web sémantique repose en effet sur des principes et des standards spécifiques : c'est ce que l'on appelle le « Web de données » ou *Linked Data*, qui est défini comme « *A web of things in the world, described by data on the Web* » (Bizer, Heath et Berners-Lee 2009). Une telle structure est rendue possible par le RDF, modèle de graphe développé par le W3C[30], qui est utilisé pour décrire les ressources sur le Web et leurs métadonnées. Avec le RDF, chaque information est exprimée sous la forme d'une assertion composée de trois éléments (ou « triplet ») constitué d'un sujet, d'un prédicat (ou propriété) et d'un nom. Le sujet et le prédicat sont représentés sous la forme d'une URI ou *Uniform Resource Identifier*, devenant alors des « ressources ». Ces ressources sont elles-mêmes subdivisées en « classes », que l'on peut définir comme des collections d'individus ou d'objets. Le Web de données n'impose pas de définition quant aux classes et aux propriétés ; la grande flexibilité du RDF n'impose pas non plus de règles d'utilisation des classes et des propriétés (Bermès 2013). Il est donc nécessaire d'adopter une technologie de représentation des connaissances qui, comme l'explique Patrick Le Bœuf, « formalis[e] la sémantique interne de l'information engrangée dans des bases qui peuvent être très hétérogènes » (Le Boeuf, Le modèle conceptuel de référence du CIDOC : de la sémantique des inventaires aux musées en dialogue 2013). C'est là qu'intervient l'ontologie : celle-ci permet de déclarer les classes, les propriétés, leurs comportements et leur hiérarchie dans des langages de représentation des connaissances comme RDF Schema (RDFS) et OWL. Choisir, au sein d'une ontologie, les classes et propriétés nécessaires pour représenter les informations que l'on souhaite traiter, permet de donner naissance à un modèle de données. Ce modèle de données consiste en l'ensemble des classes choisies pour exprimer ces informations (Bermès 2013).

[27] Le catalogue du musée est consultable à cette adresse : http://imatex.cdmt.es/.
[28] Les collections du *Victoria & Albert Museum* sont accessibles ici : https://www.vam.ac.uk/collections.
[29] On trouvera une définition du damas dans le thésaurus SILKNOW : https://skosmos.silknow.org/thesaurus/fr/page/168.
[30] https://www.w3.org/RDF/

Dans le cadre du projet SILKNOW, la création du graphe de connaissances repose sur un modèle de données basé sur un sous-ensemble de l'ontologie CIDOC *Conceptual Reference Model*. Ce sous-ensemble est instancié au moment de la conversion RDF (Schleider, Troncy et Gaitan, et al. 2021). Une ontologie est un modèle conceptuel, consensuel et partagé, qui offre le moyen de représenter toutes les informations d'un domaine d'application spécifique au moyen de concepts définis et hiérarchisés (Gruber 1993) (Guarino 1997). L'utilisation d'ontologies résout donc le problème de l'hétérogénéité sémantique posé par l'utilisation de sources de données variées (Vasilecas, Kalibatiene et Trinkunas 2006) et permet d'augmenter l'interopérabilité des systèmes et des applications (Zhuge, Xing et Shi 2008). La croissance du Web sémantique a ainsi entraîné le développement de plus en plus ontologies à des fins diverses (Wang, Xia et Niu 2014). Dans le domaine culturel et patrimonial, des ontologies de référence ont été développées pour décrire les objets culturels et modéliser les relations entre eux. La plus connue est certainement le CIDOC *Conceptual Reference Model* ou CIDOC CRM, qui est utilisé pour décrire les objets patrimoniaux conservés par les musées (Doerr, The CIDOC Conceptual Reference Model : An ontological approach to semantic interopérability of metadata 2003). L'ontologie FRBR (*Functional Requirements for Bibliographic Records*) est utilisée pour les descriptions bibliographiques (Le Boeuf, Functional Requirements for Bibliographic Records (FRBR) : Hype Or Cure-All ? 2013). Enfin FRBRoo est une « évolution orientée objet » (Kergosien, Severo et Chaudiron 2015) qui fusionne CIDOC CRM et FRBR, de façon à pouvoir modéliser à la fois les métadonnées bibliographiques et muséales (Doerr, Le Boeuf et Bekiari, FRBRoo, a conceptual model for performing arts. In Conference proceedings « The Digital Curation of Cultural Heritage 2008). Quand une ontologie existe pour un domaine, il est possible de créer des sous-ensembles ou des spécialisations de cette ontologie, correspondant à un sous-domaine spécifique (Messaoudi, et al. 2019) (Kergosien, Eric, et al. 2019).

# Le modèle de données SILKNOW

## Pourquoi le CIDOC-CRM ?

Pour créer le modèle de données SILKNOW, nous avons donc choisi d'utiliser le CIDOC CRM. Ce modèle conceptuel nous a paru tout à fait approprié pour créer le modèle de données SILKNOW, car il a été spécifiquement créé pour exprimer la sémantique sous-jacente de la documentation sur le patrimoine culturel. Le CIDOC CRM est développé et maintenu par le Comité International pour la documentation (CIDOC) de l'International Council of Museum (ICOM). Reconnu tant par le monde des musées que par celui de l'informatique, le CIDOC CRM est également une norme ISO pour la version 5.0.4 depuis 2006, norme qui a été renouvelée en 2014[31]. C'est donc un standard, qui permet d'assurer l'interopérabilité des données patrimoniales dans le cadre du Web sémantique. De nombreux projets ont ainsi utilisé des modèles de données reposant sur le CIDOC CRM pour exprimer la sémantique complexe de leurs données et les diffuser en ligne. Nous avons déjà cité les projets COURAGE (Faraj et Micsik 2021), WarSampo (Koho, et al. 2019), ArchOnto (Koch, Ribeiro et Lopes 2020) ou encore le projet travaillant sur la Compagnie Néerlandaise des Indes orientales (Schouten, et al. 2021). On pourra également citer l'alignement des données archivistiques des archives nationales portugaises avec le CIDOC CRM (Melo, Pimenta Rodrigues et Varagnolo 2021), le projet MMM (*Mapping Manuscript Migrations*) qui a pour but de pour publier trois bases de données de manuscrits hétérogènes (Hyvönen, et al. 2021), le projet MémoMines qui a développé un modèle ontologique pour représenter les connaissances du bassin minier des Hauts-de-France (Kergosien, Eric, et al. 2019), ou encore le projet Simplicius portant sur des données orales (du Château, et al. 2020). Le projet TECTONIQ a particulièrement retenu notre attention car il utilise le CIDOC CRM pour traiter des données hétérogènes ayant trait au patrimoine textile (Kergosien, Severo et Chaudiron 2015). Mais contrairement à SILKNOW, TECTONIQ s'intéresse au patrimoine industriel, composé des biens matériels (bâtiments, machines, usines, centres de productions) et des biens immatériels (souvenirs, événements, image collective, gestes). TECTONIQ ne traite donc pas les données qui décrivent les produits issu de l'activité du tissage.

Le CIDOC CRM n'est pas un standard figé : au moment de l'écriture de cet article, ce modèle conceptuel en est à la version 7.2, publiée en octobre 2021[32]. Nous utilisons la version 6.2 (Le Boeuf, Doerr, et al. 2015), alors la dernière version publiée au moment du début du projet SILKNOW en mai 2018. Le CIDOC CRM a également pour caractéristique d'être un modèle flexible et extensible. Cela signifie qu'il peut être étendu avec de nouvelles classes et de nouvelles propriétés, de façon à pouvoir exprimer de nouveaux types d'informations, sans modifier la structure de base du modèle. Il existe ainsi un certain nombre d'extensions spécialisées. Nous avons déjà cité l'extension FRBRoo utilisée pour exprimer le processus de création, de production et d'expression dans la littérature et les arts du spectacle ; il existe aussi CRMtex[33], modèle conceptuel pour les textes anciens, ou encore CRMba[34] et CRMarchaeo[35] pour la recherche en archéologie. En tout, la documentation officielle du CIDOC CRM reconnaît dix modèles compatibles[36].

Concrètement, le CIDOC CRM nous permet de modéliser de manière homogène des données par nature hétérogènes. Si nous reprenons les notices provenant du *Museu Tèxtil Terrassa* et du *Victoria & Albert Museum* (figure 6), on peut exprimer, de la même manière, les informations concernant la technique et les matériaux utilisés lors du tissage. Ces données sont alors exprimées sous la forme de triplet Classe (C) – Propriété (P) – Classe (C). Sur la figure 7, on peut voir ces triplets représentés sous la forme d'un graphe dirigé. Ce graphe est dirigé car la relation (ou propriété) entre le sujet et le prédicat (catégorisés dans des classes) est unidirectionnelle. Il faut également noter que nous avons choisi d'utiliser la classe `E22 Man-Made Object` pour représenter l'objet décrit par les métadonnées muséales. En effet, la version 6.2 du CIDOC CRM utilise cette classe pour modéliser les « *physical objects purposely created by human activity* »[37] (Le Boeuf, Doerr, et al. 2015).

[31] https://www.iso.org/standard/57832.html
[32] https://cidoc-crm.org/Version/version-7.2
[33] https://cidoc-crm.org/crmtex/ModelVersion/version-1.0-0
[34] https://cidoc-crm.org/crmba/
[35] https://cidoc-crm.org/crmarchaeo/
[36] https://cidoc-crm.org/collaborations

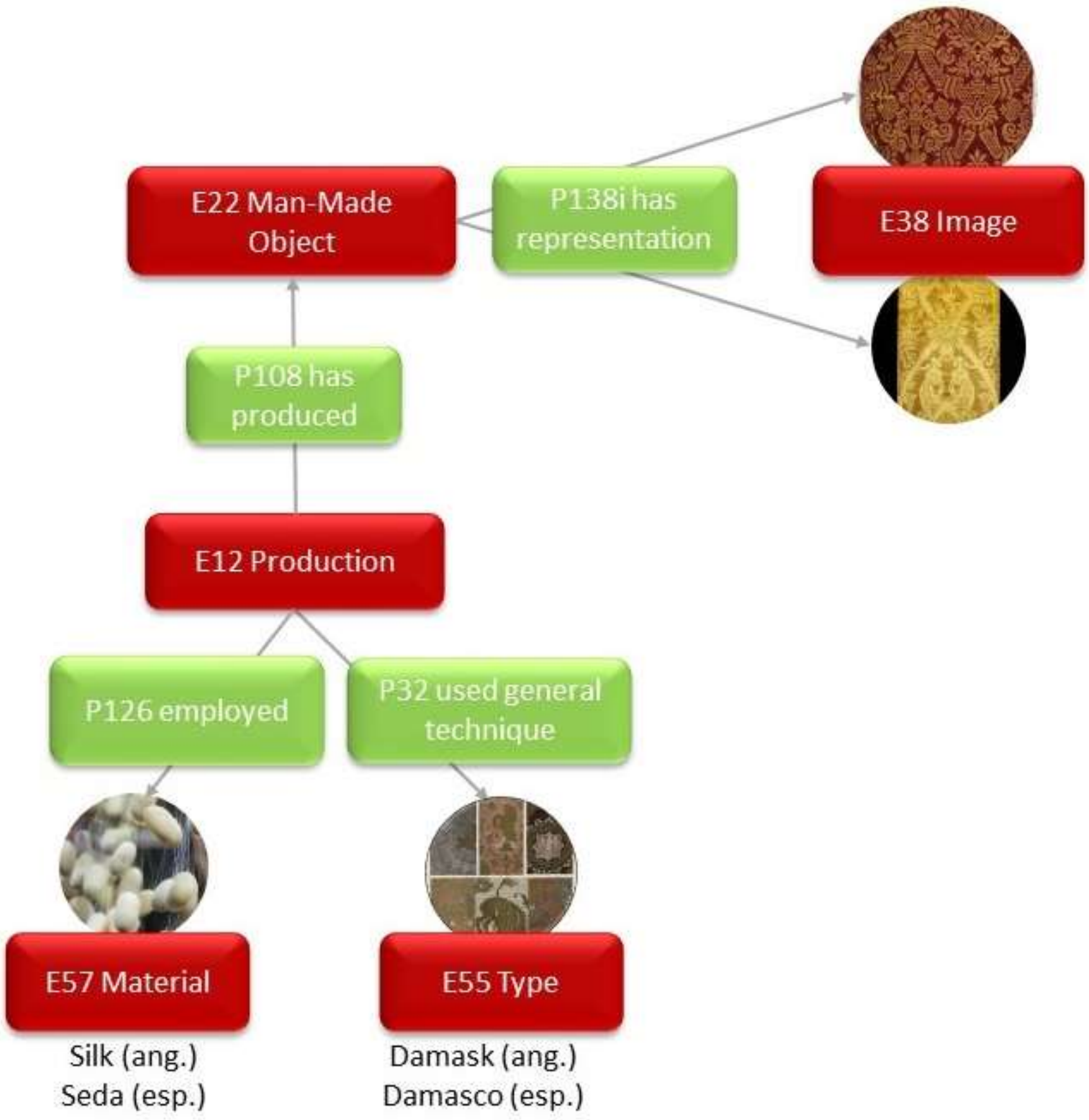

Figure 7. Graphe dirigé permettant de visualiser les triplets C-P-C utilisés pour modéliser les informations concernant la technique de tissage et les matériaux utilisés (Image produite par les auteurs)

## La sélection des classes et propriétés

Le CIDOC CRM est un modèle conceptuel ; cela signifie qu'il doit pouvoir être utilisé pour modéliser des données patrimoniales provenant de toutes sortes d'institutions et décrivant toute la gamme des objets conservés dans leurs collections. La version 6.2 du CIDOC CRM fournit ainsi 89 classes et 153 propriétés[38] ; mais il est fort peu probable qu'un modèle de données ait besoin de toutes ces classes et propriétés pour exprimer la sémantique des données qu'il est censé modéliser. Pour créer un modèle de données adapté aux métadonnées collectées par SILKNOW, il faut donc sélectionner les classes et propriétés pertinentes. Pour ce faire, il est d'abord nécessaire d'analyser en détails les données que nous voulons aligner avec le CIDOC CRM. Nous avons ainsi examiné et comparé des notices provenant des institutions patrimoniales sélectionnées, de façon à pouvoir dresser la liste des champs descriptifs qu'elles utilisent dans leurs métadonnées. Nous avons également étudié le contenu de ces champs pour en comprendre la sémantique sous-jacente et en saisir la cohérence interne. Les noms des champs descriptifs peuvent en effet être trompeurs ou bien pas assez révélateurs. Nous avons également supposé que des champs descriptifs équivalents pouvaient ne pas avoir été utilisés de la même manière par deux institutions, et que leur utilisation pouvait également avoir évolué au fil du temps. Afin de mieux comprendre le sens des champs descriptifs utilisées – autrement dit, les types d'informations exprimées -, nous nous sommes appuyés sur les standards et la documentation utilisés par ces institutions pour produire ces métadonnées, notamment les inventaires des musées français (Arrêté du 25 mai 2004 fixant les normes techniques relatives à la tenue de l'inventaire, du registre des biens déposés dans un musée de France et au récolement 2004), le modèle harmonisé HADOC pour la production de données culturelles (Briatte 2012), les directives de l'ICOM pour l'information sur les objets de musées (Grant, Nieuwenhuis et Petersen 1995) et le modèle de données *Europeana* (Europeana Data Model s.d.).

---

[37] Dans la dernière version du CIDOC CRM (Version 7.2) parue en septembre 2021, le nom de la classe E22 a évolué pour devenir « E22 Human-Made Object ».
[38] La liste de ces classes et propriétés est accessible via OntoME : https://ontome.net/namespace/1

Cette première étape nous a permis de dresser une liste des champs descriptifs les plus couramment utilisés par les institutions patrimoniales pour décrire les textiles qu'elles conservent, tout en éliminant les champs qui ne présentaient pas d'intérêt pour le projet SILKNOW. Par exemple, nous n'avons pas retenu les informations concernant la gestion administrative de ces objets. Ces champs descriptifs ont ensuite été regroupés dans des « *Information groups* » (groupes d'informations). En tout, nous avons défini 22 groupes d'informations qui forment le dictionnaire des données ou « *Data Dictionnary* » du projet SILKNOW. Un dictionnaire des données consiste en « *a collection of names, definitions, and attributes about data elements that are being used or captured in a database, information system, or part of a research project. It describes the meanings and purposes of data elements within the context of a project, and provides guidance on interpretation, accepted meanings and representation*[39] ». Comme le montre le tableau 1, nous avons par exemple défini un « *Object Acquisition Information Group* » (groupe d'informations concernant l'acquisition d'objets) pour identifier la méthode et la date d'acquisition, le précédent propriétaire de l'objet, son propriétaire actuel, et toute information supplémentaire concernant l'acquisition (notamment s'il s'agit d'un don, d'un dépôt ou d'un achat).

| **Object Acquisition and Legal Status Information Group** |
| --- |
| Definition: information about the acquisition and ownership of a cultural heritage object. Several such information groups can be available for one object depending on the history of the object.<br><br>*Acquisition method*<br>The method by which an object was acquired.<br>Ex: gift; purchase<br><br>*Acquisition time-span*<br>The timespan or the date of acquisition of the object.<br>ex: Before 1998; 1950<br><br>*Previous owner*<br>The name of the person from whom, or organization from which, the object was acquired.<br><br>*New owner*<br>The name of the person who, or organization that, acquired the object.<br><br>*Acquisition complement*<br>Any additional information about the acquisition of the object.<br><br>*Acquisition note*<br>If necessary, additional comment on the acquisition of the object |

Tableau 1. Contenu du groupe d'information « *Object acquisition and legal status information group* »

[39] Nous utilisons ici la définition donnée par la bibliothèque de l'Université de Californie que l'on trouvera à cette adresse : https://library.ucmerced.edu/data-dictionaries

Si un dictionnaire des données est particulière utile pour l'humain, il n'en va pas de même pour la machine (Rashid, et al. 2017) : il est donc nécessaire d'exprimer ce dictionnaire des données dans un langage compréhensible par un ordinateur. Les « groupes d'information » nous ont été particulièrement utiles dans cette tâche, car ils nous ont permis d'utiliser au mieux les « *Functional Overviews* » ou « aperçus fonctionnels » fournis par la documentation officielle du CIDOC CRM[40]. Ces aperçus fonctionnels divisent les classes et propriétés du modèle conceptuel en différentes catégories d'information, accompagnées de leur représentation graphique sous forme de graphes dirigés. Ces aperçus fonctionnels offrent ainsi des modèles de modélisation techniquement neutres, appliqués aux métadonnées décrivant les objets patrimoniaux. On trouve par exemple dans les aperçus fonctionnels la catégorie « *Acquisition Information* » (figure 8), qui correspond à notre catégorie « *Object acquisition and legal status information group* ».

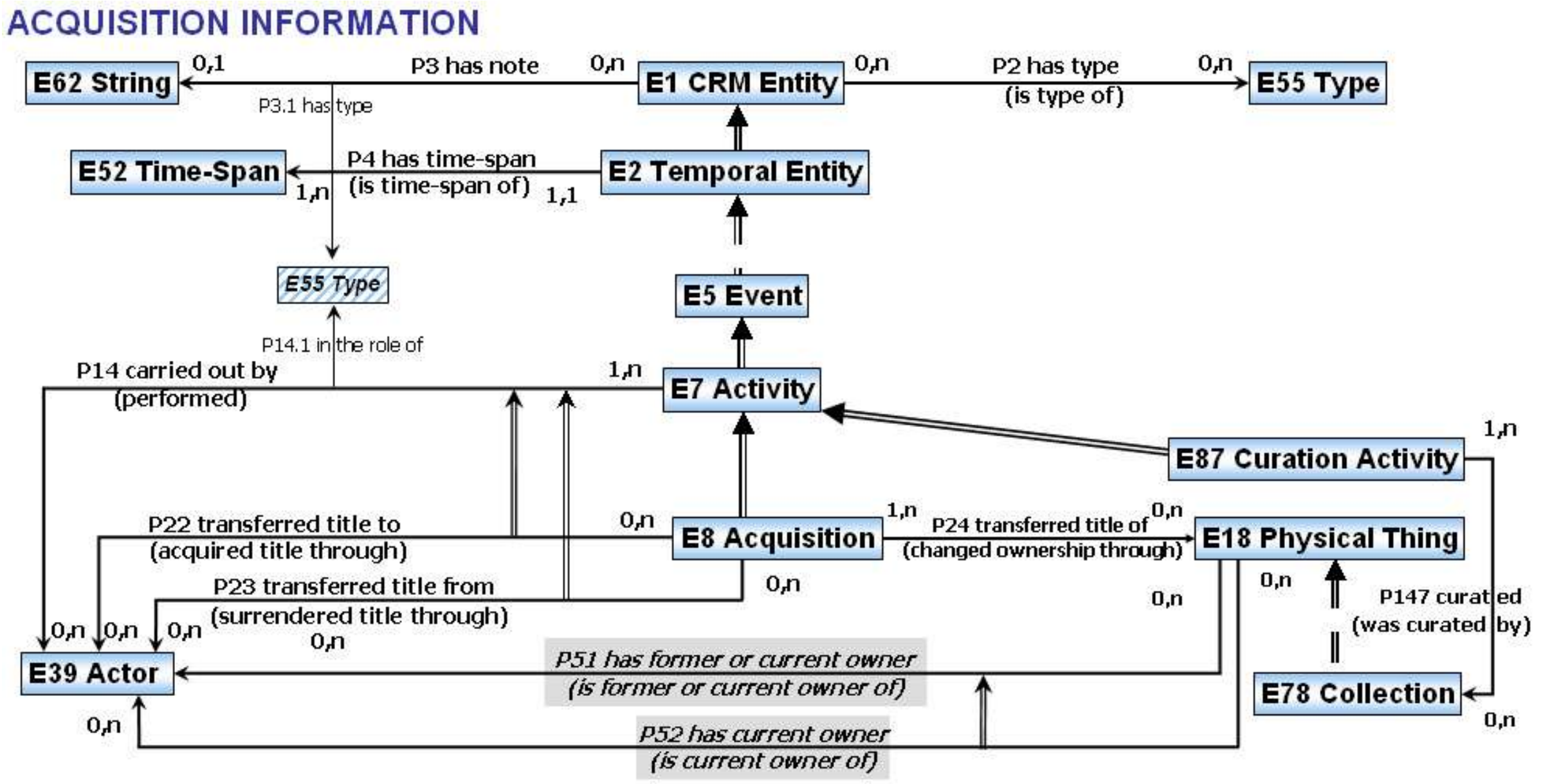

Figure 8. Graphe dirigé représentant les classes et propriétés proposées par les aperçus fonctionnels pour modéliser les informations sur l'acquisition (Source : https://cidoc-crm.org/FunctionalUnits/acquisition-information)

Les groupes d'information définis auparavant nous permettent de nous appuyer sur ces aperçus fonctionnels pour en exprimer la sémantique avec les classes et propriétés du CIDOC CRM. Dans le tableau 2, on peut voir que les classes et propriétés choisies pour exprimer le groupe d'informations concernant l'acquisition sont celles représentées dans le graphe sur la figure 8 :

| **Domain** | **Property** | **Range**[41] |
|---|---|---|
| E8 Acquisition | P14 carried out by | E39 Actor |
| E8 Acquisition | P22 transferred title to | E39 Actor |
| E8 Acquisition | P23 transferred title from | E39 Actor |
| E8 Acquisition | P24 transferred title | E22 Man-Made Object |
| E8 Acquisition | P7 took place | E53 Place |
| E8 Acquisition | P4 has time-span | E52 Time-Span |

[40] http://www.cidoc-crm.org/functional-units

[41] Les propriétés « Domain » et « Range » explicitent comme une « Property » lie un sujet et un objet dans le triplet.

Tableau 2. Classes et propriétés fournies par le CIDOC CRM et utilisées pour exprimer des informations concernant l'acquisition d'un objet par une institution patrimoniale

Sur la base de la catégorisation des données effectuée précédemment, nous avons non seulement sélectionné les classes et propriétés les plus susceptibles d'exprimer la sémantique de ces informations, mais nous avons également affiné cette première sélection en ajoutant de nouvelles classes et propriétés, et en supprimant celles qui se sont avérées, finalement, inutiles. Le modèle de données SILKNOW se composent donc des classes et propriétés sélectionnées pour exprimer la sémantique des groupes d'information que nous avons définis au préalable. La liste de ces classes et propriétés est accessible en ligne (Puren et Vernus, Silknow Generic Profile 2021) et documentée grâce au gestionnaire d'ontologie OntoME (Beretta 2021), développé et maintenu par le LARHRA, et dans lequel la documentation CIDOC CRM a été importée.

En tout, nous avons sélectionné 74 classes et 22 propriétés fournies par le CIDOC CRM Version 6.2. Nous avons également sélectionné celles offertes par une extension de ce modèle, le *Scientific Observation Model* (CRMsci) (Kritsotaki, et al. 2017). Le CRMsci est une ontologie élaborée pour intégrer les métadonnées issues de l'observation scientifique. Nous avons plus particulièrement utilisé la classe `S4 Observation`, définie comme « *the activity of gaining scientific knowledge about particular states of physical reality gained by empirical evidence, experiments and by measurements* » (Kritsotaki, et al. 2017). Cette classe nous a semblé tout à fait appropriée pour modéliser les analyses historiques (comme sur la figure 9) et les analyses techniques résultant de l'observation des tissus anciens.

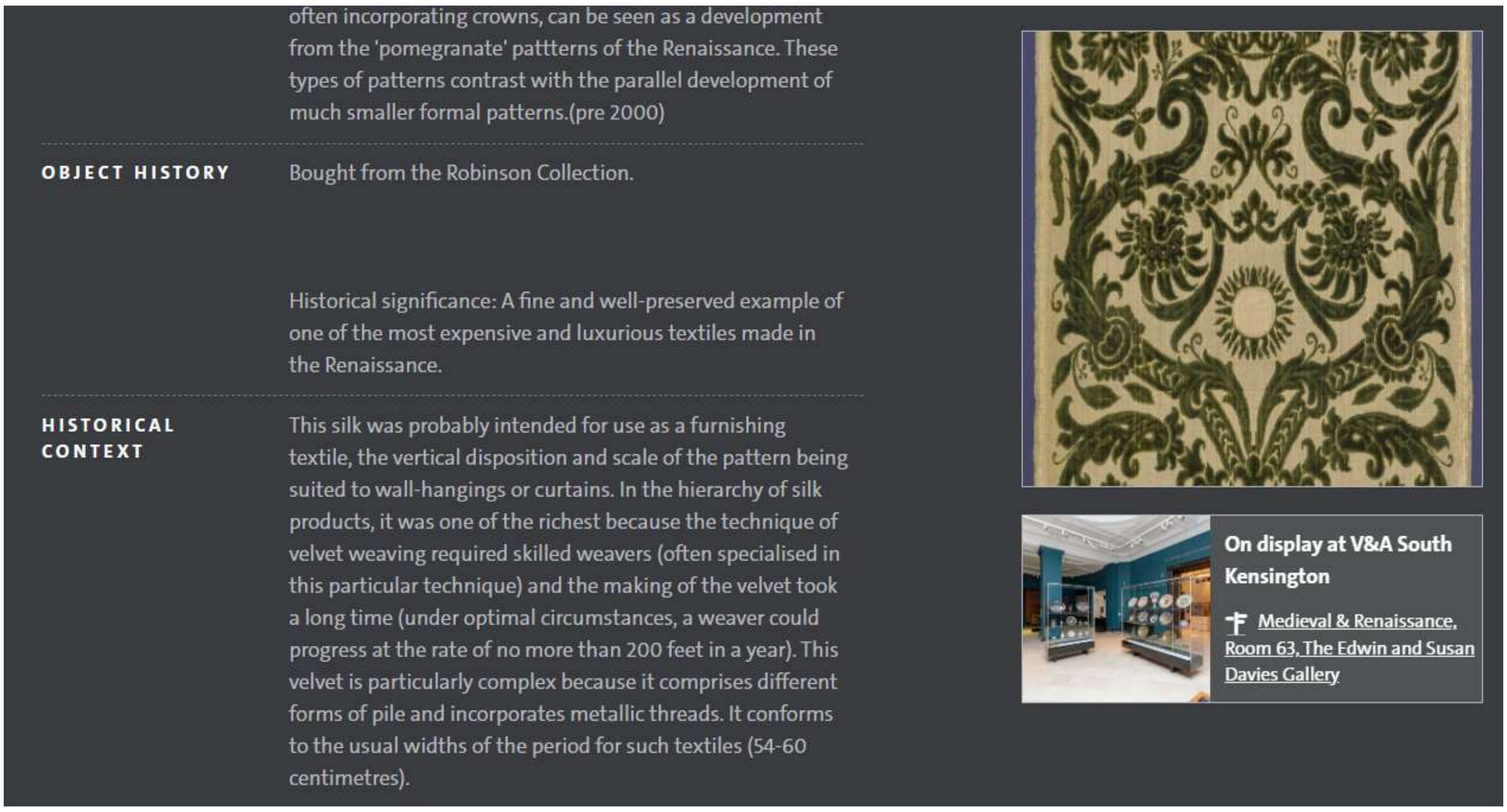

Figure 9. Exemple d'analyse historique (« *Historical Context* ») provenant du catalogue du *Victoria & Albert Museum* (Silk Velvet Furnishing Fabric, XVIe siècle[42]) (Image produite par les auteurs)

Dans l'exemple ci-dessous (figure 10), la notice provenant du catalogue des collections du Château de Versailles[43] comporte un champ « Historique » et un champ « Commentaires » correspondant respectivement à une analyse historique et à une analyse technique du tissu (ici, un lé de tenture). En utilisant le modèle de données SILKNOW, nous exprimons donc ces informations avec les triplets suivants : `S4 Observation O8 observed E22 Man-Made Object` ; `S4 Observation P2 has type E55 Type (Historique / Commentaire)` ; `S4 Observation P3 has note E62 String`.

---

[42] https://collections.vam.ac.uk/item/O13600/silk-velvet-furnishing-unknown/
[43] La notice peut être consultée à cette adresse : http://collections.chateauversailles.fr/#e7fe900c-4795-42f5-b5aa-26f2b3f649e3

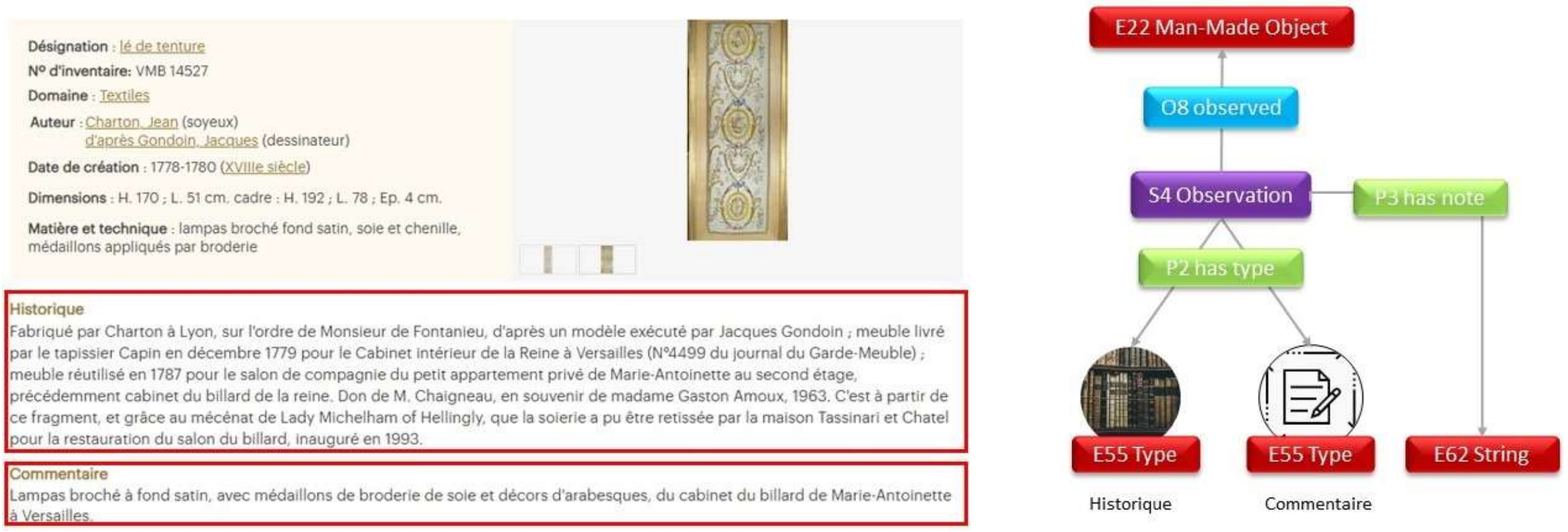

*Figure 10. Modélisation des champs « Historique » et « Commentaires » avec le modèle de données SILKNOW (CIDOC CRM et CRMsci)*[44] (Image produite par les auteurs)

## Evaluation du modèle de données et création des *mapping rules*

Cette première étape est suivie par le *mapping* ou « mise en correspondance »[45] manuelle des données. Il s'agit de mettre en équivalence des modèles de données décrivant des objets similaires (Christen 2012) : ici nous mettons en équivalence les modèles de données utilisés par les institutions patrimoniales avec le modèle de données SILKNOW. Le *mapping* consiste à produire des données sémantiques à partir des métadonnées produites par les institutions patrimoniales, qui sont stockées dans des bases de données relationnelles. Il s'agit plus exactement de donner une expression sémantique équivalente à chacun des champs descriptifs au moyen des classes et propriétés fournies par le modèle de données. On fournit alors des *mapping rules* ou règles de mise en correspondance. Le *mapping* a d'une part pour fonction de vérifier la qualité et la robustesse du modèle de données, et d'autre part de permettre la conversion des fichiers de métadonnées en RDF grâce aux règles ainsi définies.

La mise en correspondance a été réalisée manuellement par des spécialistes du domaine formés aux humanités numériques. Concrètement, pour chacune des institutions sélectionnées, nous avons choisi deux objets décrits de manière extensive dans les catalogues. Pour cette étape, nous avons utilisé les fichiers des métadonnées collectés, convertis et stockés dans un format JSON intermédiaire. Sur la figure 11, on peut voir un extrait du fichier JSON correspondant à la notice ci-dessus (figure 10), et qui comporte cinq champs descriptifs :

---

[44] Les classes issues du CIDOC CRM sont représentées en rouge et les propriétés en vert ; les classes provenant du CRMsci sont en violet et les propriétés en bleu turquoise.

[45] Parfois traduit par « mappage » (Office Québécois de la langue français, « Mise en correspondance », *Grand Dictionnaire Terminologique*, 2001,
URL : http://gdt.oqlf.gouv.qc.ca/ficheOqlf.aspx?Id_Fiche=8873847)

```
"fields": [
    {
        "label": "title",
        "value": "lé de tenture"
    },
    {
        "label": "Désignation",
        "value": "lé de tenture"
    },
    {
        "label": "N° d'inventaire:",
        "value": "VMB 14527"
    },
    {
        "label": "Domaine",
        "value": "Textiles"
    },
    {
        "label": "Date de création",
        "value": "XVIIIe siècle"
```

Figure 11. Extrait du fichier de métadonnées converti dans le format JSON, décrivant le lé de tenture conservé dans les collections du Château de Versailles (Image produite par les auteurs)

Nous avons ensuite interprété chacun des champs descriptifs comme une entité-relation-entité (e-r-e), selon la méthode suggérée par Haridimos Kondylakis, Martin Doerr et Dimitri Plexousakis (Kondylakis1, Doerr et Plexousakis 2006). Plus précisément,

- les tables et les colonnes de la base de données relationnelle sont interprétées comme des entités,
- les notices complètes sont interprétées comme des instances d'entités,
- les noms des champs sont interprétés à la fois comme des relations et des entités,
- et le contenu des champs est interprété comme des instances d'entités.

L'ensemble du schéma est décomposé en e-r-e, et chaque e-r-e est aligné sur le CIDOC CRM (Doerr, Mapping a Data Structure to the CIDOC Conceptual Reference Model 2003). En d'autres termes, le *mapping* consiste à interpréter ces entités et relations, et à les exprimer dans la sémantique du CIDOC CRM. Ce faisant, nous cherchons à préserver autant que possible le sens original des données. Ce processus produit des triplets qui relient des nœuds (les classes) entre eux par des liens (les propriétés), formant un réseau de données lisibles par l'homme et la machine et permettant l'échange et l'intégration des informations.

Le tableau 3 montre le résultat de cette mise en correspondance pour les cinq champs représentés sur la figure 11. Dans cet exemple, le champ « Title » contient le titre donné par l'institution patrimoniale. Nous exprimons la sémantique sous-jacente de ce champ descriptif de la manière suivante : le titre de l'artefact est « Lé de tenture » dans la base de données. Cela signifie que nous pouvons interpréter ce champ comme un titre, modélisé par la classe `E35 Titre`. Le nom du champ décrit la relation qui existe entre l'objet `E22 Man-Made Object` et son titre, ce qui implique de l'interpréter avec la propriété `P102 has title`. On remarque également que les champs « Title » et « Désignation » contiennent la même valeur « Lé de tenture ». Toutefois nous ne les avons pas interprétés de la même manière. « Title » correspond clairement au titre donné à l'objet, tandis que « Désignation » a pour vocation d'exprimer le type d'objet dont il s'agit. Le champ « Domaine » a le même rôle, mais la catégorie est ici beaucoup plus large. Avec le triplet `E17 Type Assignment P14 carried out by E40 Legal Body (Château de Versailles)`, nous avons également exprimé le fait que le contenu de ces champs était le résultat d'une décision prise par l'institution conservant l'objet, à la suite d'une analyse de celui-ci.

| **Nom du champ descriptif** | **Contenu** | **Triplets exprimés avec les classes et propriétés provenant du CIDOC CRM** |
|---|---|---|
| Title | Lé de tenture | `E22 Man-Made Object P102 has title E35 Title` |
| Désignation | Lé de tenture | `E17 Type Assignment P41 classified E22 Man-Made Object`<br>`E17 Type Assignment P42 assigned E55 Type (Lé de tenture)`<br>`E17 Type Assignment P14 carried out by E40 Legal Body (Château de Versailles)`<br>`E17 Type Assignment P2 has type E55 Type (Désignation)` |
| N° d'inventaire | VMB 14527 | `E15 Identifier Assignment P140 assigned attribute to E22 Man-Made Object`<br>`E15 Identifier Assignment P37 assigned E42 Identifier`<br>`E15 Identifier Assignment P14 carried out by E40 Legal Body (Château de Versailles)`<br>`E42 Identifier P2 has type E55 Type (N° d'inventaire)` |
| Domaine | Textiles | `E17 Type Assignment P41 classified E22 Man-Made Object`<br>`E17 Type Assignment P42 assigned E55 Type (Textiles)`<br>`E17 Type Assignment P14 carried out by E40_Legal Body (Château de Versailles)`<br>`E17 Type Assignment P2 has type E55 Type (Domaine)` |
| Date de création | XVIIIe siècle | `E12 Production P4 has time-span E52 Time-Span`<br>`E52 Time-Span P78 is identified by E49 Time Appellation (XVIIIe siècle)` |

Tableau 3. Mise en correspondance des données contenues dans le fichier JSON avec les classes et propriétés provenant du modèle de données SILKNOW

Nous avons créé des règles de mise en correspondance pour chacune des notices sélectionnées, accompagnées de leurs représentations sous forme de graphes dirigés. Ceux-ci nous permettent de visualiser les triplets – parfois complexes - et de nous assurer ainsi de leur cohérence. Le graphe ci-dessous (figure 12) représente les triplets créés pour modéliser les métadonnées provenant de la notice représentée sur la figure 10 – notamment ceux que nous avons systématiquement utilisés pour exprimer les conditions de production d'un tissu. Le triplet `E12 Production P108 has produced`

`E22 Man Made Object` est ainsi accompagné des triplets suivants, pour exprimer les informations concernant :

- la date de production : `E12 Production P4 has time span E52 Time-Span` ;
- le lieu de production : `E12 Production P8 took place on or within E53 Place` ;
- son(ses) créateur(s)/ producteur(s) (quand l'information est disponible) : `E12 Production P14 carried out by par E39 Actor` ;
- le(s) matériau(x) utilisé(s) : `E12 Production P126 employed E57 Material` ;
- la(es) technique(s) utilisée(s) : `E12 Production P32 use general technique E55 Type`.

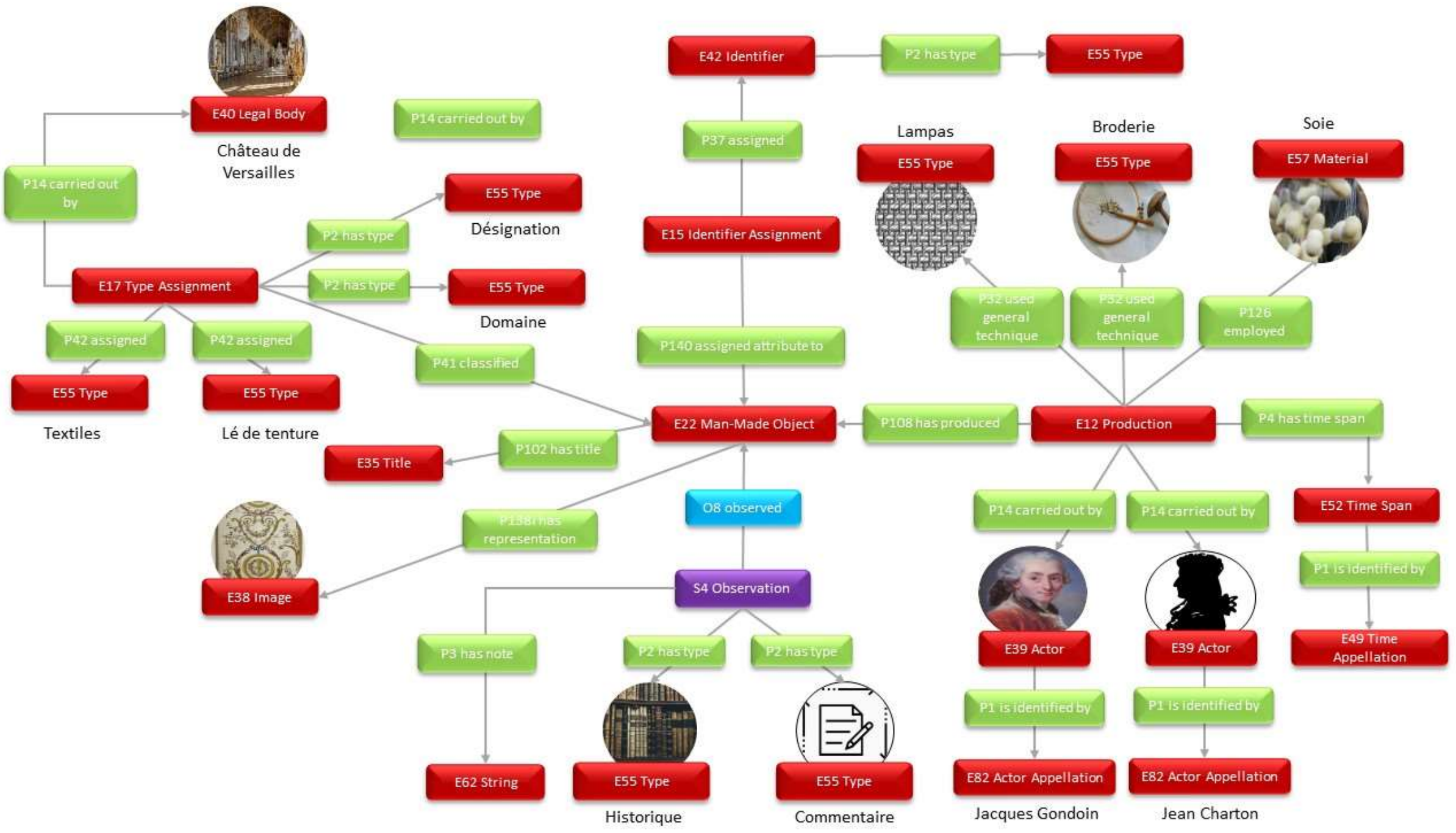

Figure 12. Graphe dirigé représentant une partie des triplets utilisés pour modéliser les métadonnées de la notice provenant du catalogue des collections du Château de Versailles (Image produite par les auteurs)

On remarquera l'ajout du triplet `E22 Man-Made Objet P138i has representation E38 Image`. Il n'existe pas de champ « Image » dans les métadonnées, mais celui-ci existe de manière implicite. Il est en effet nécessaire de modéliser la relation existant entre l'objet et sa représentation, de manière à les relier aux données décrivant les objets correspondants.

Nous avons observé que tous les champs descriptifs pouvaient être représentés en utilisant les classes et propriétés sélectionnées dans le modèle de données. Les fichiers JSON ont alors été convertis au moyen d'un convertisseur RDF[46] pour créer le graphe de connaissances. La conversion produit des fichiers Turtle (Terse RDF Triple Language) (Schleider, Troncy et Ehrhart, et al. 2021) (Schleider, Troncy et Gaitan, et al. 2021), qui est l'une des syntaxes utilisées pour exprimer des triplets RDF. Les syntaxes RDF sont conçues pour partager, exposer et diffuser des triplets RDF (Bermès 2013). Le graphe obtenu contient en tout 36210 objets illustrés par 74527 images. Les fichiers Turtle sont ensuite chargés dans un triple store basé sur le serveur Virtuoso, et les images sont téléchargées séparément sur un serveur multimédia (Schleider, Troncy et Ehrhart, et al. 2021) (Schleider, Troncy et Gaitan, et al. 2021). Virtuoso est un triple store open source : c'est une base de données développée spécifiquement pour stocker et manipuler les données RDF, qui sont exprimées comme des « sujets », « prédicats » et

[46] https://github.com/silknow/converter

« objets » (Bermès 2013). Les données contenues dans le graphe RDF sont accessibles via une interface de requête SPARQL[47] (Schleider, Troncy et Ehrhart, et al. 2021) (Schleider, Troncy et Gaitan, et al. 2021). SPARQL est le langage et le protocole de requête pour le RDF développé par le W3C (Bermès 2013). Il est également possible d'explorer ces données via un navigateur à facettes[48]. SILKNOW a également conçu une interface graphique pour explorer ces données : il s'agit du moteur de recherche exploratoire ADASilk (*Advanced Data Analysis for Silk Heritage*)[49] (Schleider, Troncy et Ehrhart, et al. 2021).

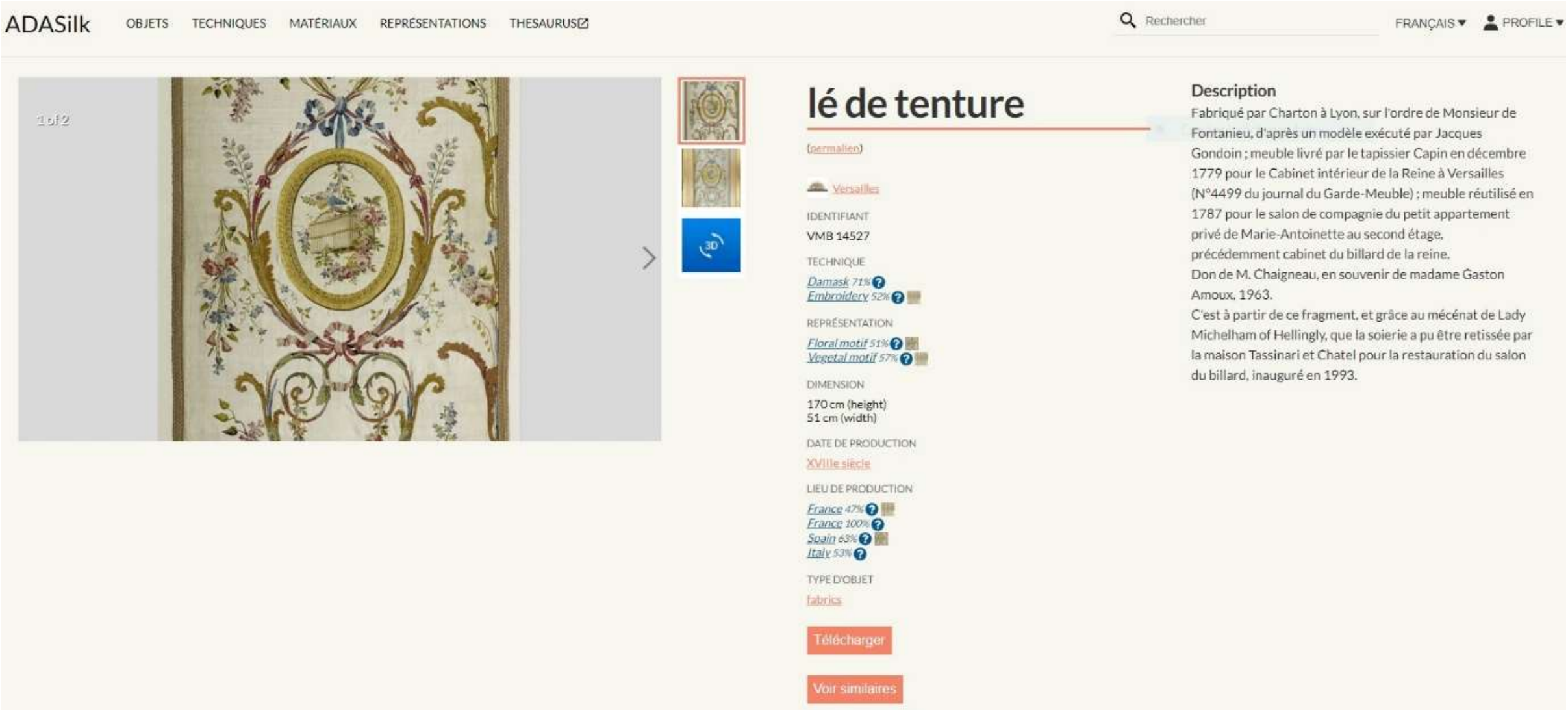

Figure 13. Visualisation dans ADASilk des métadonnées décrivant le « lé de tenture » conservé dans les collections du Château de Versailles (Image produite par les auteurs)

# Enrichissement des données et intégration des annotations dans le modèle de données

## Utilisation de Prov DM : *Provenance Data Model*

SILKNOW a également développé des méthodes d'analyse de textes (Massri et Mladenić 2020) (Schleider et Troncy, Zero-Shot Information Extraction to Enhance a Knowledge Graph Describing Silk Textiles 2021) et d'images (Dorozynski et Clermont, Muti-task deep learning with incomplete training samples for the image-based prediction of variables describing silk fabrics 2019). Basées sur l'intelligence artificielle, celles-ci à infèrent de nouvelles propriétés sur les objets à partir des métadonnées collectées. Ces analyses ont donc pour objectif de produire de nouvelles données sur ces objets et d'enrichir les métadonnées existantes ; ces annotations sont également accessibles en ligne via ADASilk. Dans la notice représentée à la figure 6, on trouve dans le champ « Technique » les informations suivantes : « Damask 71% » et « Embroidery 52% » ; et dans le champ « Représentation » : on lit « Floral motif 51% » et « Vegetal motif 57% » (cf. figure 14). Il s'agit des prédictions réalisées à partir des métadonnées associées à l'objet décrit.

[47] https://data.silknow.org/sparql
[48] https://data.silknow.org/fct/
[49] https://ada.silknow.org/fr

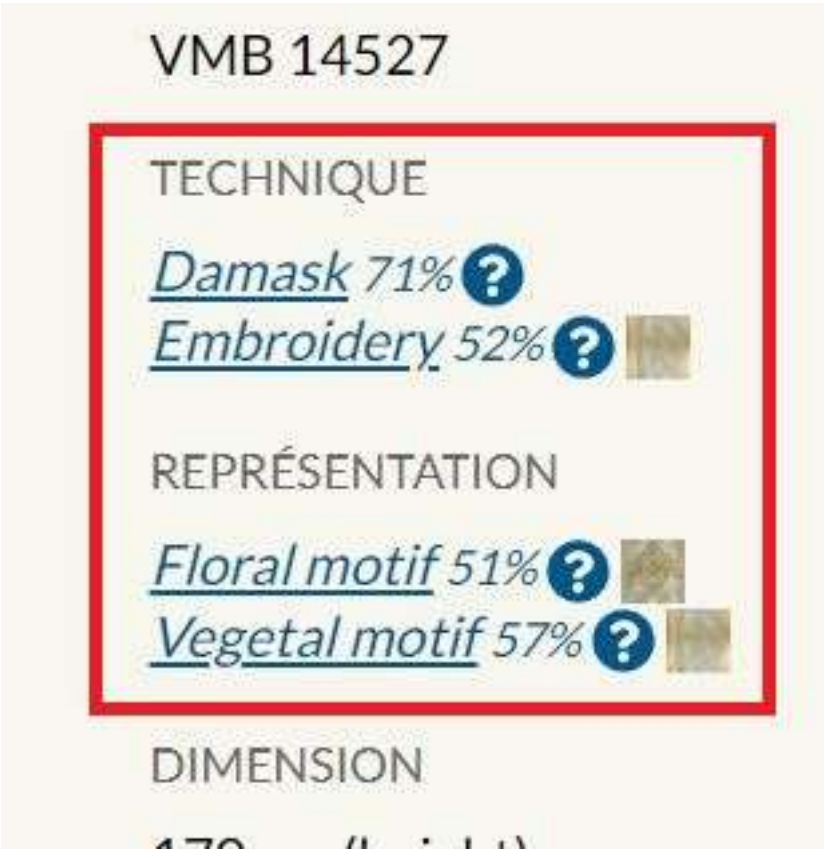

Figure 14. Prédictions réalisées sur un tissu à partir des métadonnées (textes et images), accompagnées de leurs scores de confiance (Image produite par les auteurs)

Ces annotations ont été obtenues en utilisant des réseaux de neurones convolutionnels entraînés sur les données (textes et images) contenues dans le graphe de connaissances. Avec en entrée les textes décrivant les objets ou les images les représentant, des modèles de classification des textes (Rei 2021) ou d'images (Dorozynski et Rottensteiner, SILKNOW Image Classification - final version WP6 - Multi-Task Classification (1.0) 2021) peuvent inférer la valeur de certaines propriétés, à savoir : la période de production, le lieu de production, la technique et le matériau pour le modèle de classification de textes ; pour le modèle de classification d'images, les mêmes propriétés auxquels s'ajoutent les motifs représentés,.

Pour intégrer ces nouvelles informations dans le graphe de connaissances, il est nécessaire de modéliser ces données en ajoutant de nouvelles classes et propriétés au modèle de données. La modélisation doit permettre de faire une distinction claire entre ces prédictions et les données originales, et fournir aux utilisateurs d'ADASilk des informations sur le degré de fiabilité de ces informations. Pour cela, nous avons choisi d'utiliser le *Provenance Data Model* (Prov DM) (Belhajjame, et al. 2013), recommandé par le W3C. Prov DM est un modèle conceptuel qui permet d'exprimer les informations concernant la provenance d'une donnée ou d'une chose. Le graphe ci-dessous (figure 15) représente les triplets que nous avons utilisés pour intégrer ces annotations avec Prov DM dans le modèle de données[50]. Les analyses d'images ou de textes sont représentées sous la forme d'une `Prov:Activity` qui peut être qualifiée par un type (analyse d'image ou analyse de texte). Selon le cas, cette `Prov:Activity` prend une `E38 Image` ou un texte `E62 String` en entrée (`prov:used`) et produit deux déclarations en sortie (propriétés `prov:WasGeneratedBy`). Chacune de ces déclarations possède une `E54 Dimension`, permettant d'exprimer le score de confiance associé. La date de l'analyse peut également être précisée (`prov:AtTime`). Si nécessaire, on peut spécifier le module d'analyse avec une classe `prov:Agent` (de type *Software Agent*) et le documenter (`E31 Document`).

[50] Les métadonnées exprimées sous formes de triplets sur le graphe dirigé décrivent un lé conservé au *Musée des Arts décoratifs* : https://ada.silknow.org/fr/object/99528c86-aac9-3231-a9d9-b84f6e4756fd.

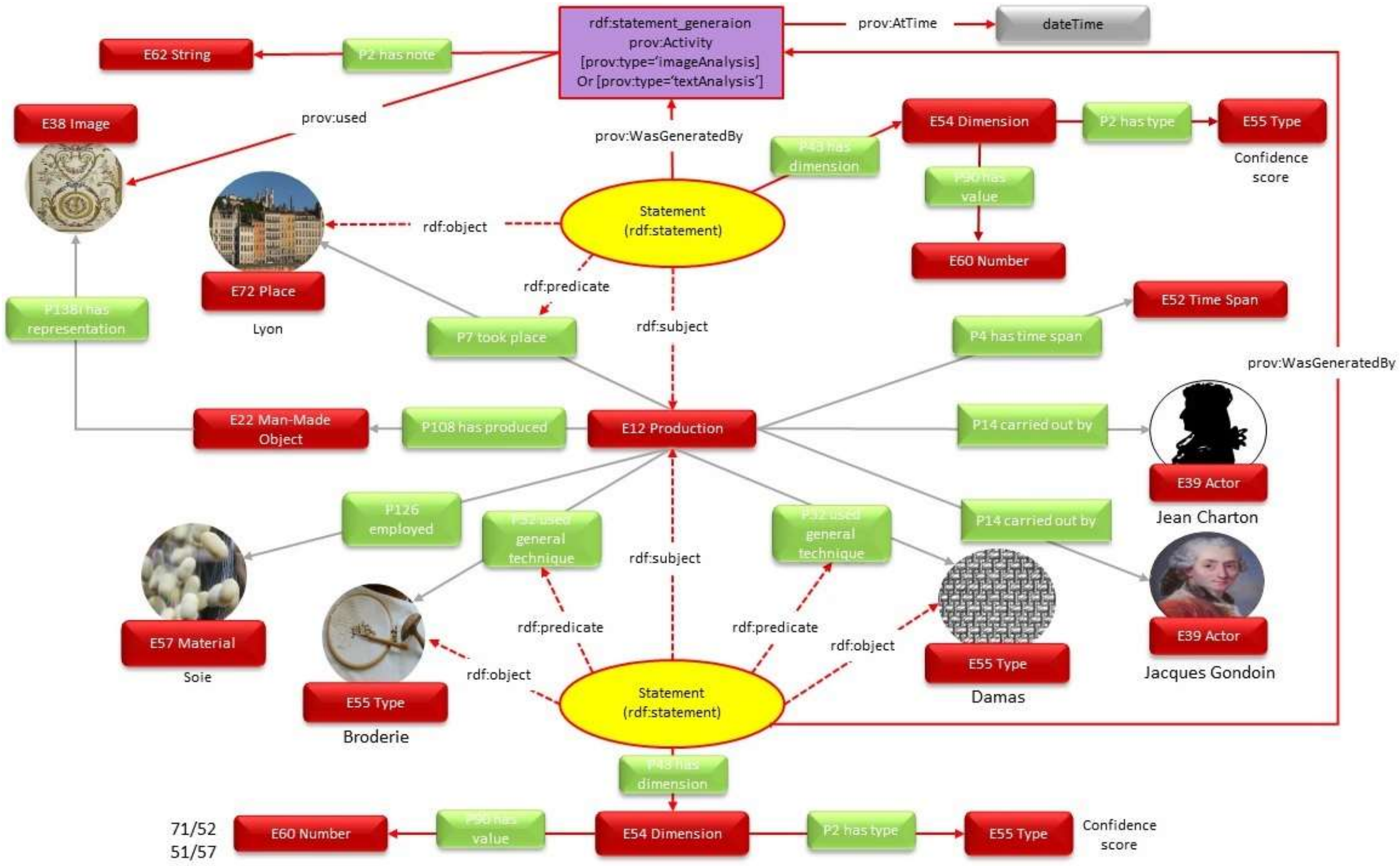

Figure 15. Modélisation des données issues de l'analyse d'images avec ProvDM (Image produite par les auteurs)

# Proposition d'une extension pour l'analyse technique des tissus

On a vu que du texte libre était souvent utilisé, dans les métadonnées, pour rédiger des analyses techniques et historiques sur les tissus. Ces textes, parfois longs, ont été rédigés par des historiens de l'art et par des experts de la fabrication textile, et ils fournissent fréquemment des informations particulièrement intéressantes sur le processus de tissage. L'analyse de la structure et de la décoration des tissus, et la présentation du contexte de leur production et de leur utilisation, sont ainsi exprimées dans ces blocs de textes. Les analyses techniques, menées en amont du catalogage, permettent ainsi de déterminer quelle(s) armure(s)[51] ont été employées ou encore quel(s) type(s) de fil(s) a(ont) été utilisé(s). Le CIDOC CRM a toutefois des difficultés à capturer la richesse sémantique de ces textes libres. Par ailleurs, le *mapping* que nous avons réalisé vise avant tout à stocker ces métadonnées « telles quelles », car nous souhaitions intégrer rapidement ces données dans le graphe de connaissances. Ces informations ont donc été stockées sous forme de « note » en utilisant le CIDOC CRM et le CRMsci, au moyen du triplet `S4 Observation P3 has note E62 String` (cf. figure 10). La propriété `P3 has note` fonctionnant comme « *a container for all informal descriptions about an object that have not been expressed in terms of CRM constructs* » (Le Boeuf, Doerr, et al. 2015), elle peut être utilisée pour modéliser toutes les informations caractérisant un objet, quel que soit la nature de cette caractérisation. Mais cette propriété ne permet pas d'exprimer de manière structurée « *everything that can be said about an item* » (Le Boeuf, Doerr, et al. 2015). C'est une des conséquences du formalisme du CIDOC CRM qui n'a pas pour vocation à exprimer « tout ce qui peut être dit » sur un objet.

[51] Le terme « armure » désigne ici la manière dont les fils de trame et les fils de chaîne ont été entrecroisés pour créer un tissu. On parle par exemple d'« armure satin ». Une armure est utilisée pour produire un effet particulier et pour créer des motifs. Dans les tissus complexes, on peut utiliser plusieurs armures différentes. Pour en savoir plus sur ce terme, on pourra consulter le thésaurus SILKNOW : https://skosmos.silknow.org/thesaurus/fr/page/318.

L'extraction d'informations à partir de ces données textuelles montre à quel point ces observations produisent des analyses techniques détaillées, et offrent de nouvelles perspectives historiques sur les tissus anciens. Par exemple, ces blocs de texte peuvent analyser les techniques de tissage utilisées, décrire les motifs représentés, ou encore resituer le tissu au sein d'un style particulier. Comme l'illustre la figure 15, l'analyse menée sur ces blocs de texte a ainsi permis d'en extraire, de manière systématique, les informations suivantes : les armures, les techniques de tissage, les motifs et les styles.

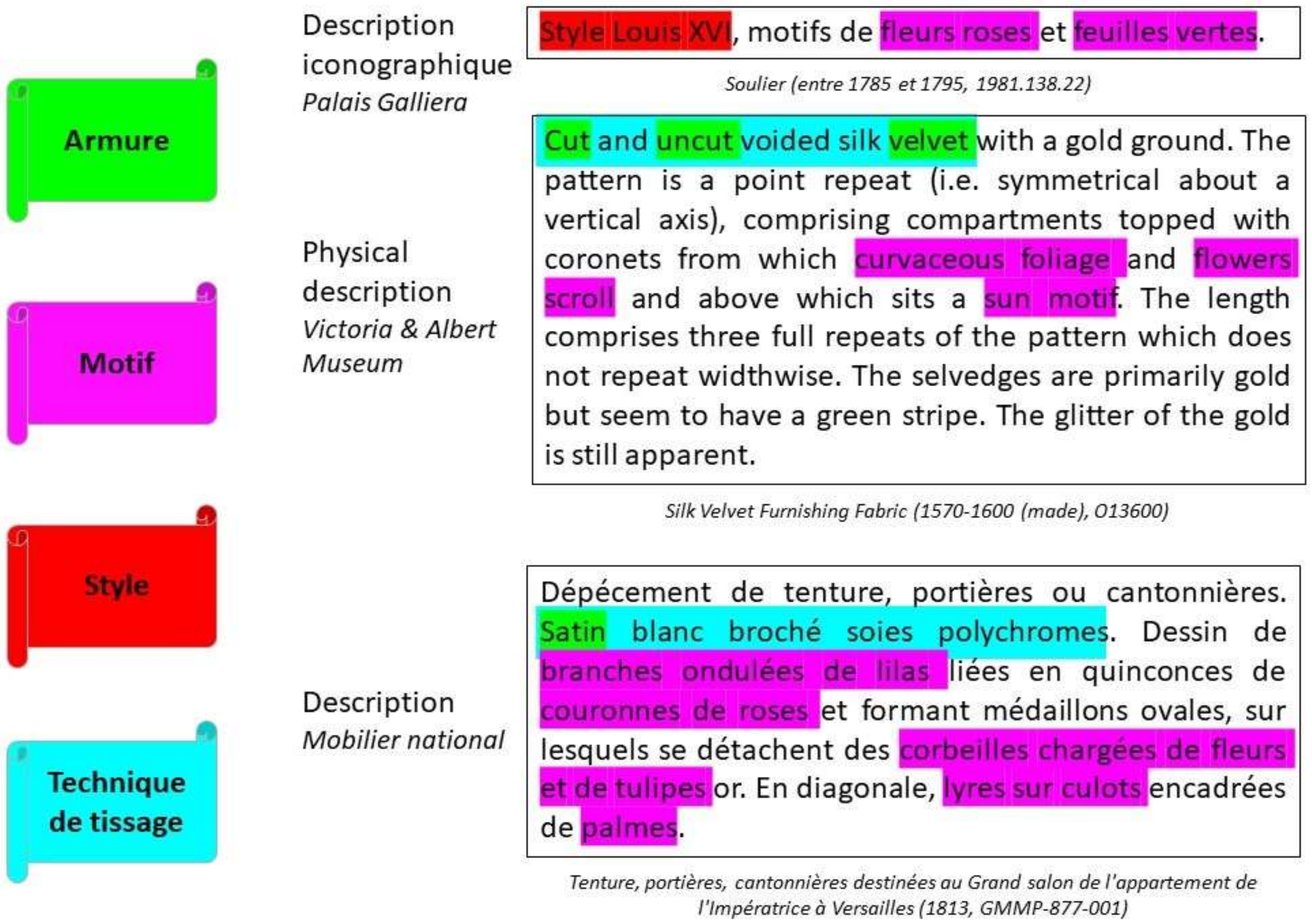

Figure 16. Les quatre types d'informations extraits des descriptions à partir d'exemples provenant du *Victoria & Albert Museum* et des collections du *Mobilier national* (Image produite par les auteurs)

En choisissant d'exprimer ces informations par une simple note, celles-ci ne sont pas directement représentées dans le graphe de connaissances, ce qui n'en facilitera pas l'accès via un moteur de recherche, ni le traitement automatique. Les utilisateurs ne seront pas en mesure de formuler des requêtes fines sur ces données. Pourtant, ces informations peuvent intéresser des publics variés : le grand public pourrait explorer les notices en s'intéressant à des motifs spécifiques ; les historiens des techniques pourraient retracer plus facilement l'usage de certaines technologies ; les industries créatives pourraient s'inspirer de styles historiques.

Il faut également noter qu'au moment de l'intégration des données dans le graphe de connaissances, les informations concernant les techniques et les motifs sont annotées grâce au thésaurus multilingue SILKNOW, consacré au vocabulaire des textiles anciens en soie[52] (Léon, et al. 2019). Le processus d'enrichissement sémantique est réalisé de la manière suivante : la valeur d'une chaîne de caractères dans les métadonnées est remplacée par un identifiant unique, lorsque que celle-ci coïncide avec la valeur d'une chaîne de caractères correspondant à une technique de tissage ou à un motif dans le thésaurus (Schleider, Troncy et Gaitan, et al. 2021). En modélisant ces concepts avec des classes et des propriétés appartenant au modèle de données, il serait donc possible d'intégrer ces annotations dans le graphe de connaissances, de donner accès à ces nouvelles données via ADASilk, et de créer ainsi des liens entre ces concepts et le thésaurus – et dans le cadre du Web sémantique, avec d'autres données ouvertes liées. Il s'agit ici de donner véritablement corps aux objectifs du Web sémantique, en créant des liens avec d'autres données et en enrichissant l'expérience utilisateur.

[52] Disponible à cette adresse, en français, anglais, italien et espagnol : https://skosmos.silknow.org/thesaurus/

On a vu que le CIDOC CRM ne fournissait pas de classes et de propriétés permettant d'exprimer la sémantique de ces informations. Il est toutefois possible d'exploiter la flexibilité inhérente à ce modèle conceptuel, qui permet de développer des extensions plus spécialisées – comme le *Scientific Observation Model* par exemple, ou FRBRoo. Nous n'avons pas l'intention de capturer « tout ce qui peut être dit » sur les objets que nous étudions, mais d'exprimer des informations spécifiques à leur sujet. Dans le cadre du projet SILKNOW, nous nous intéressons plus particulièrement au processus ayant abouti à la production d'un tissu ancien en soie. En utilisant les résultats de l'extraction des données et l'annotation de ces dernières, nous avons donc créé un modèle compatible avec le CIDOC CRM destiné à exprimer de manière formelle ce processus. Cette extension spécialisée est accessible en ligne et documentée dans OntoME, et elle propose 23 classes et 12 propriétés (Puren et Vernus, SIKNOW 0.1. 2021).

Pour créer cette extension, nous avons adopté une approche ascendante, en nous basant avant tout sur l'analyse des données collectées. Nous avons également travaillé en collaboration avec des experts du domaine et des informaticiens pour vérifier l'intérêt et la validité de ces nouvelles classes et propriétés. Nous avons notamment créé la classe `T1 Weaving` (sous-classe de `E12 Production`) qui désigne l'activité consistant à entrecroiser à angles droits, en général à l'aide d'un métier à tisser, les fils de chaîne (les fils dans la longueur du tissu, tendus sur le métier à tisser), et les fils de trame (les fils perpendiculaires qui passent alternativement entre les fils de chaîne)[53].

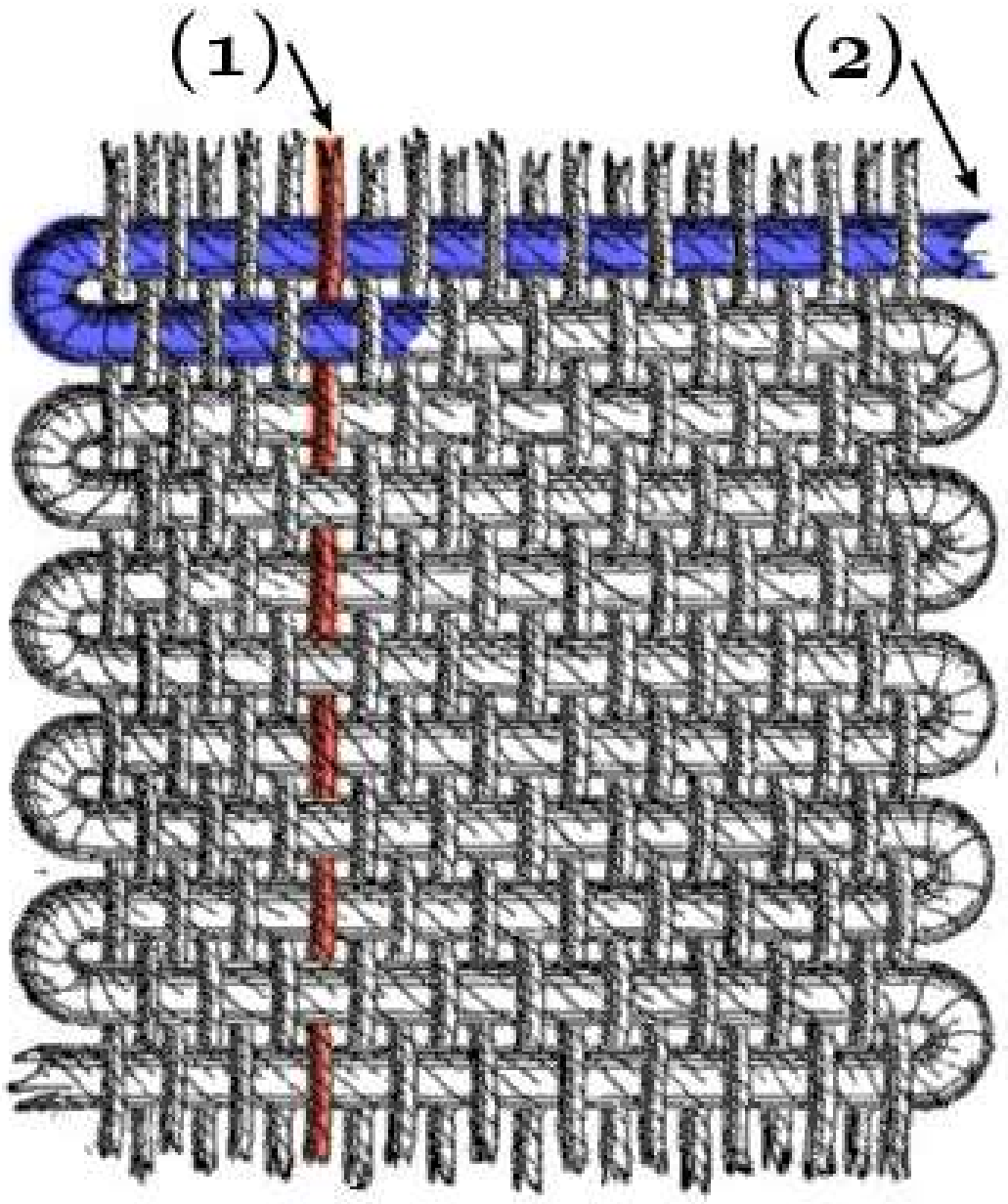

Figure 17. Figure représentant les fils de chaîne (1) et les fils de trame (2) (Source : https://fr.wikipedia.org/wiki/Tissage#/media/Fichier:Kette_und_Schu%C3%9F_num_col.png, licence CC-BY-SA)

Le tissage permet de produire un tissu ou `T7 Fabric` (sous-classe de `E22 Man-Made Object`). Ces deux entités peuvent être reliés grâce aux propriétés `P108 has produced` (`T1 Weaving P108 has produced T7 Fabric`) et `P31 has modified` (`T1 Weaving P31 has` modified T7 Fabric). La production du tissu est réalisée en suivant une ou plusieur(s) procédures techniques (T25 Weaving Technique, sous-classe de E29 Design or Procedure). Le tissage exige également d'utiliser une armure (T21 Weave, sous-classe de E55 Type) spécifique (T32 Weave Type, sous-classe de E55 Type). Le tissu est la plus plupart du temps décoré de motifs (T18 Motif, sous-classe de E22 Man-Made Object) de différentes sortes (T34 Motif Type, sous-classe de E55 Type), apparaissant sur une zone délimitée du tissu (T9 Pattern Zone, sous-classe de E22 Man-Made Object). Ces motifs sont souvent caractéristiques d'un style (T11 Style, sous-classe de E55 Type) identifié par des experts du domaine (T13 Style Assignment, sous-classe de E13 Attribute Assignment). Le graphe ci-dessous (figure 18) montrent comment utiliser ces nouvelles classes et propriétés pour exprimer les informations fournies par les analyses techniques et historiques. Nous avons utilisé ici un exemple provenant des collections du Château de Versailles[54].

[53] Pour plus d'informations sur le tissage : https://skosmos.silknow.org/thesaurus/fr/page/526

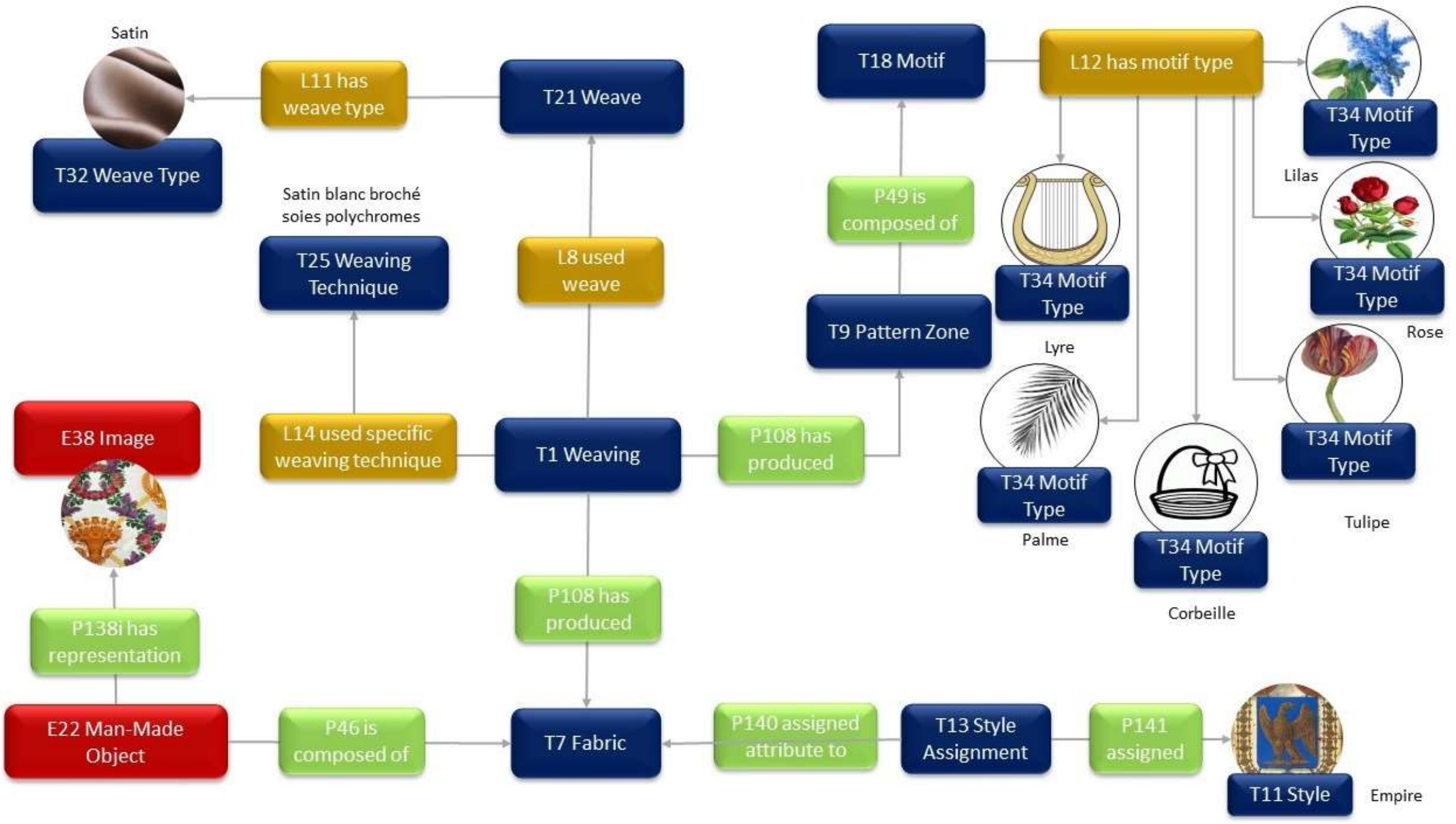

Figure 18. Graphe dirigé conforme au modèle conceptuel CIDOC CRM et à l'extension SILKNOW[55] ((Image produite par les auteurs)

Ces nouvelles classes et propriétés doivent permettre de faciliter l'intégration des enrichissements sémantiques dans le modèle de données et dans le graphe de connaissances, et donc de permettre à l'utilisateur final d'interroger ces annotations et d'obtenir de nouveaux éléments de contextualisation. Par exemple, les classes T32 Weave Type et T34 Motif Type ont pour vocation de modéliser les informations concernant respectivement le(s) différent(s) type(s) d'armure(s) utilisé(s) durant le tissage, et le(s) différent(s) motif(s) résultant de cette activité. Comme le montre la figure 18, on peut exprimer ces informations en utilisant les classes et propriétés fournies par l'extension SILKNOW : T21 Weave L11 has weave type T32 Weave Type et T18 Motif L12 has motif type T34 Motif Type. Ces classes et propriétés sont également destinés à créer des liens entre ces informations et les définitions disponibles dans le thésaurus SILKNOW (cf. figure

---

[54] « Tenture, portières, cantonnières destinées au Grand salon de l'appartement de l'Impératrice à Versailles » (1813, Bissardon, Cousin & Bony), Collections du Mobilier national, URL : https://collection.mobiliernational.culture.gouv.fr/objet/GMMP-877-001

[55] Les nouvelles classes sont en bleu foncé, et les nouvelles propriétés en jaune moutarde.

19), permettant à l'utilisateur de consulter facilement la définition de l'armure « Satin » dans le thésaurus[56] ou encore, celle du motif « Palme »[57].

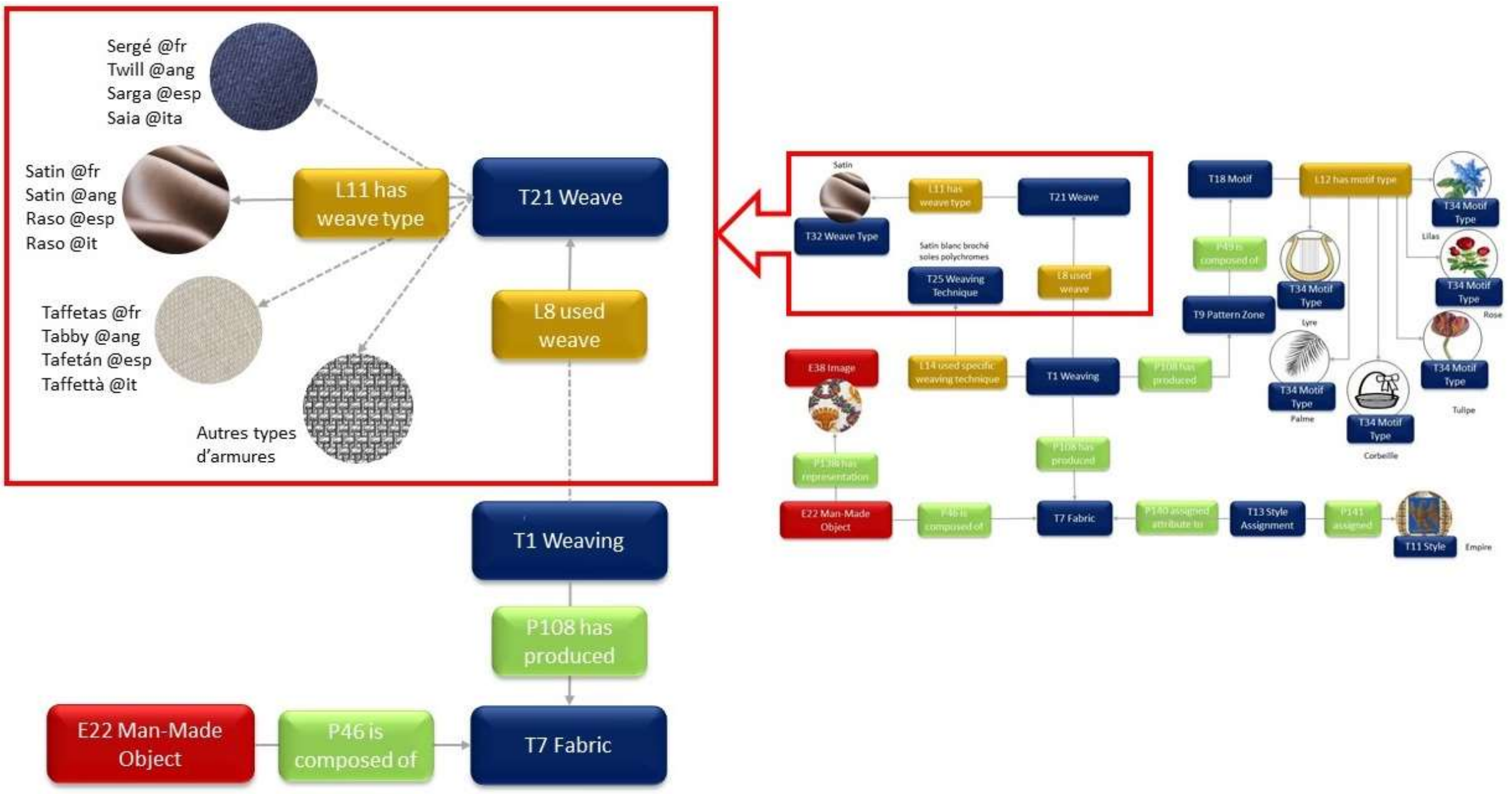

Figure 19. Utilisation de la classe `T32 Weave Type` (Image produite par les auteurs)

# Conclusion

SILKNOW avait pour but de mieux valoriser le patrimoine soyeux européen. En créant le moteur de recherche exploratoire ADASilk, le projet a montré l'intérêt d'utiliser les technologies du Web sémantique pour agréger des données patrimoniales hétérogènes, y donner accès et accroître leur visibilité. Il s'inscrit ainsi dans la lignée d'autres projets ayant également pour but de valoriser des patrimoines fragiles et menacés, et il montre la validité de ce type d'approches pour le traitement à grande échelle des données patrimoniales. SILKNOW illustre également l'important travail à réaliser sur les données patrimoniales pour donner véritablement corps aux objectifs du Web sémantique. Si les institutions patrimoniales ont compris l'intérêt de « libérer » leurs données et de s'assurer de leur interopérabilité, il reste encore du chemin à faire pour qu'elles produisent systématiquement des données ouvertes liées ou « données 5 étoiles », respectant les principes énoncés par Tim Berners-Lee en 2006 (Berners-Lee, Linked Data 2006).

Les ré-utilisateurs de ces données ont également un rôle crucial à jouer dans la modélisation de ces données. Ils peuvent en effet proposer des modèles de données collaboratifs et partagés pour rendre ces données interopérables et pleinement réutilisables dans le contexte du Web sémantique. L'extension SILKNOW constitue l'une des étapes qui pourraient, par exemple, mener à la création d'une ontologie de domaine pour le patrimoine textile – ontologie qui tiendrait compte non seulement des résultats de la production textile (les tissus), mais aussi des lieux et techniques de production, des conditions de cette production, et de la mémoire qui lui est associée. Plus largement, une telle approche pourrait être systématisée lorsque l'on veut valoriser et protégé un patrimoine menacé.

Si des projets tels que SILKNOW ont montré les apports des technologies sémantiques pour la protection du patrimoine, il semble également nécessaire de s'interroger sur leur facilité d'utilisation et d'application. On voit que ces projets demandent des investissements importants, et qu'ils restent donc inaccessibles pour les petites institutions qui en seraient, pourtant, les premières bénéficiaires. Après avoir démontré l'intérêt de ces approches, il faudrait donc envisager de développer des outils donnant accès à ces technologies, et les mettre ainsi à la portée du plus grand nombre.

[56] https://skosmos.silknow.org/thesaurus/fr/page/237
[57] https://skosmos.silknow.org/thesaurus/fr/page/763

# Références